\documentclass[twocolumn, prc, amssymb,  aps, showpacs,preprintnumbers,
amsmath,showkeys,floatfix]{revtex4}

\usepackage{epstopdf}
\usepackage{graphics}
\usepackage{graphicx}
\usepackage{dcolumn}
\usepackage{bm}
\usepackage{longtable}
\usepackage{epsfig}
\usepackage{times}
\usepackage{url}

\begin{document}
%%%%%%%%%%%%%%%%%%%%%%%%%%%%%%%%%%%%%%%%%%%%%%%%%%%%%%%%%%%%%%%%%%%%%%%%%%%%%%%%
%%%%%%%%%%%%%%%%%%%%%%%%%%%%%%%%%%%%%%%%%%%%%%%%%%%%%%%%%%%%%%%%%%%%%%%%%%%%%%%%

\title{ A three-beam setup for coherently controlling nuclear state population}

\author{Wen-Te \surname{Liao}}
\email{Wen-Te.Liao@mpi-hd.mpg.de}

\author{Adriana \surname{P\'alffy}}
\email{Palffy@mpi-hd.mpg.de}

\author{Christoph~H. \surname{Keitel}}
\email{Keitel@mpi-hd.mpg.de}

\affiliation{Max-Planck-Institut f\"ur Kernphysik, Saupfercheckweg 1, 69117 Heidelberg, Germany}
\date{\today}
%%%%%%%%%%%%%%%%%%%%%%%%%%%%%%%%%%%%%%%%%%%%%%%%%%%%%%%%%%%%%%%%%%%%%%%%%%%%%%%%
%%%%%%%%%%%%%%%%%%%%%%%%%%%%%%%%%%%%%%%%%%%%%%%%%%%%%%%%%%%%%%%%%%%%%%%%%%%%%%%%
\begin{abstract}
The controlled transfer of nuclear state population using two x-ray laser pulses is investigated theoretically. 
The laser pulses drive two nuclear transitions in a nuclear three-level system facilitating coherent population transfer via the quantum optics technique of stimulated Raman adiabatic passage. To overcome present limitations of the x-ray laser frequency, we envisage accelerated nuclei interacting with two copropagating or crossed x-ray laser pulses in a three-beam setup. We present a systematic study of this setup providing both pulse temporal sequence and laser pulse intensity for optimized control of the nuclear state population. The tolerance for geometrical parameters such as laser beam divergence of the three-beam setup as well as for the velocity spread of the nuclear beam are studied and a two-photon resonance condition to account for experimental uncertainties is deduced. This additional condition gives a less strict requirement for the experimental implementation of the three-beam setup. Present experimental state of the art and future prospects are discussed.

\end{abstract}
%%%%%%%%%%%%%%%%%%%%%%%%%%%%%%%%%%%%%%%%%%%%%%%%%%%%%%%%%%%%%%%%%%%%%%%%%%%%%%%%
%%%%%%%%%%%%%%%%%%%%%%%%%%%%%%%%%%%%%%%%%%%%%%%%%%%%%%%%%%%%%%%%%%%%%%%%%%%%%%%%
\maketitle
%%%%%%%%%%%%%%%%%%%%%%%%%%%%%%%%%%%%%%%%%%%%%%%%%%%%%%%%%%%%%%%%%%%%%%%%%%%%%%%%
%-----------Text body-----------------------------------------------------------

\section{Introduction} \label{sec:NCPT-intro}
%%%%%%%%%%%%%%%%%%%%%%%%%%%%%%%%%%%%%%%%%%%%%%%%%%%%%%%%%%%%%%%
Unlike the case of atoms, coherent control of nuclear states remains so far challenging \cite{baldwin81,matinyan98,kocharovskaya1999}. Typically, the traditional way of shifting nuclei from one internal quantum state to another is  by incoherent photon absorption, i.e., incoherent $\gamma$-rays (usually bremsstrahlung) illuminate the nuclear sample and excite the nuclei to some high-energy states. Subsequently, some of the excited nuclei may decay to the target state by chance, according to the corresponding branching ratio. This kind of method is rather passive, and its efficiency is low. Encouraged by the development of the X-ray Free Electron Laser (XFEL) \cite{feldhaus1997,saldin2001,arthur2002,altarelli2009,yabashi2010,lcls2}, an improved version was considered \cite{dzyublik2010} using coherent x-ray absorption. However, even this scheme stays in the traditional and passive frame and only increases somewhat the excitation efficiency. Recently, a promising setup for nuclear coherent population transfer in a three-level system using the quantum optics technique of stimulated Raman adiabatic passage  (STIRAP) \cite{bergmann1998}
has been proposed \cite{liao2011}. Two overlapping x-ray laser pulses drive two nuclear transitions and allow for coherent population transfer directly between the nuclear states of interest without loss via incoherent processes. 
This would enable actively manipulating the nuclear state by using coherent hard x-ray photons and lay an important milestone in the new developing field of nuclear quantum optics \cite{buervenich2006,wong2011,adams2013}.

\begin{figure}
\begin{center}
\includegraphics[width=8cm]{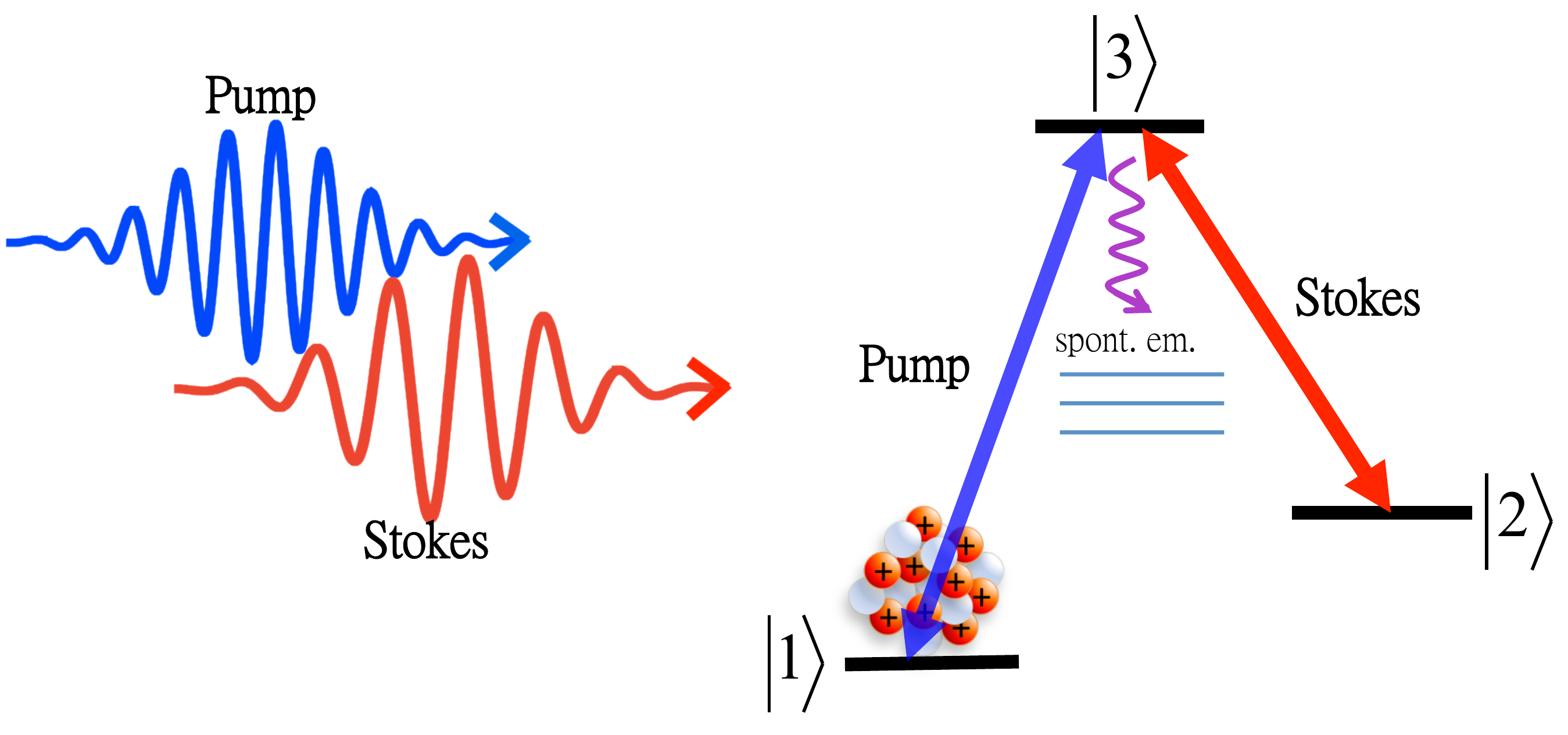}
\caption{The $\Lambda$-level scheme. The blue arrow illustrates the pump pulse, and the red arrow depicts the Stokes pulse. All initial population is in state $\vert 1\rangle$.} \label{NCPT_threelevels}
\end{center}
\end{figure}

Efficient control of the nuclear population dynamics in a three-level system as the one shown in Fig.~\ref{NCPT_threelevels} (also referred to as $\Lambda$-type system) is in particular interesting due to its association to level schemes necessary for isomer depletion. Nuclear metastable states, also known as isomers, can store large amounts of energy over longer periods of time. Isomer depletion, i.e., release on demand of the energy stored in isomers, has received a lot of attention in the last one and a half decades, especially related to the fascinating prospects of nuclear batteries \cite{walker1999,ledingham2003,palffy2007}. 
Two notable examples on triggered $\gamma$-emission from nuclear isomers by x-ray absorption are $^{180m}$Ta \cite{collins1988, belic1999, carroll2004} and $^{178m2}$Hf \cite{collins1999}, though the results reported in Ref. \cite{collins1999} remain under debate \cite{mcnabb2000,neumann2000,olariu2000,collins2000,carroll2004,carroll2009}. As shown in Fig.~\ref{NCPT_threelevels}, by shining only the pump radiation pulse, depletion occurs when the nuclear population in isomer state $|1\rangle$ is excited to a higher triggering level $|3\rangle$ whose spontaneous decay to other lower levels, e.g., state $|2\rangle$ is no longer hindered by the long-lived isomer. However, such  nuclear state control is achieved by incoherent processes (spontaneous decay) and its efficiency is therefore low. In this paper we consider the efficient coherent nuclear population transfer setup proposed in Ref.~\cite{liao2011}. Two x-ray laser pulses, the pump and the Stokes pulse, drive the two nuclear transitions $|1\rangle \rightarrow |3\rangle$ and $|2\rangle \rightarrow |3\rangle$, respectively. Since most of the nuclear transition energies are higher than the energies of the  currently available coherent x-ray photons, 
an accelerated nuclear target is envisaged, i.e., a nuclear beam produced by particle accelerators \cite{buervenich2006}. This allows for a match of the x-ray photon and nuclear transition frequency in the nuclear rest frame. This three-beam setup using the STIRAP technique has been originally proposed in Ref.~\cite{liao2011}.

Here we present an extensive study of the three-beam setup for control of nuclear population taking into account realistic parameters for experimental implementation. An optimization procedure for the required laser intensity as a function of the chosen pulse delay is presented and applied for 
 an extended case study involving a number of nuclear isotopes. While in Ref.~\cite{liao2011} ideal laser beam divergence and a monochromatic nuclear beam were considered, we address here the realistic case taking into account the available conditions at experimental facilities. We deduce a first-order two-photon resonance condition that allows to connect the error of the XFEL divergence angle with the velocity uncertainty of the nuclear beam. This additional condition gives a less strict requirement for the experimental implementation and can be used to design the parameters of the XFEL and nuclear bunches. These parameters are discussed in conjunction with the availability of both x-ray coherent lasers and ion accelerators.  In addition to the large scale infrastructures, we propose to utilize tabletop solutions, which are well developed in the field of laser-plasma interaction.  This tabletop approach also deserves attention for the problem of a global experimental availability of the nuclear state manipulation.

This paper starts with a brief overview of the STIRAP theory and the model used for the nuclear coherent population transfer calculations in Section~\ref{Theory}. The collinear and crossed beam setups are presented in Section~\ref{Setup} and the corresponding resonance conditions taking into account beam divergency and velocity spread are deduced. Our numerical results and the optimization procedure for the laser parameters are presented in Section~\ref{NumRes}. Section 
~\ref{ExpFacil} discusses the availability of experimental facilities. The paper concludes with a short Summary.

% effective pulse shape?

\section{Theoretical Approach} \label{Theory}
%%%%%%%%%%%%%%%%%%%%%%%%%%%%%%%%%%%%%%%%%%%%%%%%%%%%%%%%%%%%%%%

%%%%%%%%%%%%%%%%%%%%%%%%%%%%%%%%%%%%%%
\subsection{Master equation}

The XFEL-nuclei interaction in the nuclear rest frame is illustrated by the level scheme in Fig.~\ref{NCPT_threelevels}. The nuclear dynamics is governed by the master equation for the nuclear density matrix $\widehat{\rho}(t)$ \cite{bergmann1998, scully2006,liao2011}
\begin{equation}\label{NCPT_eq1}
\frac{\partial}{\partial t}\widehat{\rho} = \frac{1}{i\hbar}\left[ \widehat{H},\widehat{\rho}\right]+\widehat{\rho}_{s}, 
\end{equation}
with the interaction Hamiltonian
\begin{eqnarray}
\widehat{H} &=& -\frac{\hbar}{2}
\left( 
\begin{array}{ccc}
 0 & 0 & \Omega_{p}^{*}\\
  0 & -2\left(\bigtriangleup_{p}-\bigtriangleup_{S} \right)  & \Omega_{S}^{*}\\ 
  \Omega_{p} & \Omega_{S} & 2\bigtriangleup_{p}
\end{array}  
\right),
\end{eqnarray}
and the decoherence matrix
\begin{eqnarray}
\widehat{\rho}_{s} &=& \frac{\Gamma}{2}
\left( 
\begin{array}{ccc}
  2B_{31}\rho_{33} & 0 & -\rho_{13}\\
  0 & 2B_{32}\rho_{33} & -\rho_{23}\\ 
  -\rho_{31} & -\rho_{32} & -2\rho_{33}
\end{array}  
\right).
\end{eqnarray}
The initial conditions  are
\begin{equation}\label{NCPT_ic}
\rho_{ij}(0) = \delta_{i1}\delta_{1j}. 
\end{equation}
Here we consider the general case with $\theta$ denoting the angle between the pump and the Stokes laser beams. 
The angle $\theta$ is zero for the pump laser and $\theta=\theta_S$ for the Stokes laser.
The quantities appearing in the equations above are defined in Table~\ref{NCPT_table1}.
So far, Eq.~(\ref{NCPT_eq1}) - Eq.~(\ref{NCPT_ic}) are the standard approaches for any quantum system like that showed in Fig.~\ref{NCPT_threelevels}. The nuclear physics and relativitic treatment for the nuclei in the accelerated beam will enter the whole calculation when deriving the values of Rabi frequencies \cite{blatt1991,palffy2008} in the nuclear rest frame, 
\begin{eqnarray}
\Omega_{p}(t) &=& \frac{1}{\hbar}\langle 3\vert \widehat{H}_{I}\vert 1\rangle \nonumber \\
&=& \frac{4}{\hbar}\sqrt{\frac{\pi I_{p}^{\mathrm{eff}}(t)}{c \epsilon_{0}}}\sqrt{\frac{(2I_{1}+1)(L_{31}+1)}{L_{13}}}
\nonumber \\
&\times&
\frac{k^{L_{31}-1}_{31}}{(2L_{13}+1)!!}\sqrt{\mathbb{B}(\varepsilon/\mu\, L_{13})}, \\
\Omega_{S}(t) &=& \frac{1}{\hbar}\langle 3\vert \widehat{H}_{I}\vert 2\rangle \nonumber \\
&=& \frac{4}{\hbar}\sqrt{\frac{\pi I_{S}^{\mathrm{eff}}(t)}{c \epsilon_{0}}}\sqrt{\frac{(2I_{2}+1)(L_{23}+1)}{L_{23}}}
\nonumber \\
&\times&
\frac{k^{L_{23}-1}_{32}}{(2L_{23}+1)!!}\sqrt{\mathbb{B}(\varepsilon/\mu\, L_{23})},
\end{eqnarray}
where $I_{p(S)}^{\mathrm{eff}}(t)$ is the Gaussian pump (Stokes) pulse in the nuclear rest frame:
\begin{eqnarray}
I_{p(S)}^{\mathrm{eff}}(t)&=&\gamma^{2}\left( 1+\beta\cos\theta\right)^{2}
I_{p(S)}^{\mathrm{eff}}
\nonumber \\
&\times&
\mathrm{Exp}\left\lbrace -\left[\frac{\gamma(1+\beta \cos\theta)(t-\tau_{p(S)})}{T_{p(S)}}\right]^{2}\right\rbrace. 
\end{eqnarray}
Combining the expressions above, the slowly varying effective Rabi frequencies $\Omega_{p(S)}(t)$ in the nuclear rest frame for nuclear transitions of electric ($\varepsilon$) or  magnetic ($\mu$) multipolarity $L$ are given by \cite{bergmann1998,palffy2008,liao2011}
\begin{eqnarray}
\Omega_{p(S)}(t)  &=& \frac{4\sqrt{\pi}}{\hbar}\gamma^2(1+\beta \cos\theta)^2I^{\mathrm{eff}}_{p(S)}\frac{k_{31(2)}^{L_{1(2)3}-1}}{(2L_{1(2)3}+1)!!}
\nonumber \\ 
&\times&
\left[\frac{(L_{1(2)3}+1)(2I_{1(2)}+1)\mathbb{B}(\varepsilon/\mu\, L_{1(2)3})}{c\epsilon_{0}L_{1(2)3}}\right]^{1/2} 
 \nonumber \\ 
 &\times&  \mathrm{Exp}\left\lbrace -\left[\frac{\gamma(1+\beta \cos\theta)(t-\tau_{p(S)})}{\sqrt{2}T_{p(S)}}\right]^{2}\right\rbrace\, . 
\label{NCPT_eq4}
\end{eqnarray}

Here we have expressed the nuclear multipole moment  with the help of the reduced transition probabilities $\mathbb{B}(\varepsilon/\mu\, L)$  following the approach developed in  Refs.~\cite{palffy2008,liao2011}. This allows for a unified treatment of the laser-nucleus interaction for both dipole-allowed ($E1$) and dipole-forbidden nuclear transitions.  
All the laser quantities have been transformed in Eq.~(\ref{NCPT_eq4}) into the nuclear rest frame, leading to 
\begin{itemize} 
  \item the angular frequency $\gamma(1+\beta \cos\theta)\omega_{p(S)}$,
  \item bandwidth  $\gamma(1+\beta \cos\theta)\Gamma_{p(S)}$,
  \item pulse duration  $T_{p(S)}/(\gamma(1+\beta \cos\theta))$,
  \item laser peak intensity  $\gamma^{2}(1+\beta \cos\theta)^{2}I_{p(S)}$.
\end{itemize}
We emphasize that the most important parameter $\mathbb{B}(\varepsilon/\mu\, L)$ characterizing the strength of the nucleus-radiation interaction is obtained from the experimental data, e.g., the Nuclear Structure and Decay Databases \cite{nsdd}, such that no first principle calculation involving specific nuclear models is needed.

%%%%%%%%%%%%%%%%%%%%%%%%%%%%%%%%%%%%%%%%%%%%%%%%%%%%%%%%%%%%%%%%%%%%%%%%%
%\begin{widetext}
\begin{table}[t]
\centering
\caption{\label{NCPT_table1}
The notations used throughout the text. The indices $i,j=1,2,3$ denote the three nuclear states showed in Fig.~\ref{NCPT_threelevels}. The label `Lab' (`Rest') indicates that the corresponding values are in the lab (nuclear rest) frame.
}
\begin{tabular}{c|c|l}
\hline
\hline
Notation & Frame & \multicolumn{1}{c}{Explanation} \\
\hline
%\colrule
$c$ & Any & Speed of light in vacuum. \\
$\beta$ & Lab & Velocity of the nuclear particle, in units of $c$.\\
$\gamma$ & Lab & Relativistic factor $\gamma=1/\sqrt{1-\beta^{2}\cos^{2}\theta}$.\\%$\gamma=\frac{1}{\sqrt{1-\beta^{2}}}$.\\
$\epsilon_{0}$ & Any &  Vacuum permittivity. \\
$h$ & Any & Planck constant, $h=2\pi\hbar$. \\
$\Gamma$ & Rest & Spontaneous decay rate of $\vert 3 \rangle$.\\
$\rho_{ij}$ & Rest &  Density matrix element. \\
$\delta_{ij}$ & Any & Kronecker delta. \\
$k_{3i}$ & Rest & Wave number of $\vert3\rangle\rightarrow\vert i \rangle$ transition. \\
$B_{3i}$ & Rest & Branching ratio of $\vert3\rangle\rightarrow\vert i \rangle$ spontaneous decay.\\
$\Gamma_{Lp(S)}$ & Lab & Laser bandwidth of pump (Stokes). \\
$\Omega_{p(S)}$ & Rest & Slowly varying effective Rabi frequency of  \\
& & pump (Stokes) laser.\\
$\tau_{p(S)}$ & Rest & Temporal peak position of pump (Stokes) laser. \\
$\omega_{p(S)}$ & Lab & Angular frequency of pump (Stokes) laser,\\
$\Delta_{p(S)}$ & Rest & Laser detuning \\
 & & $\Delta_{p(S)}=\gamma(1+\beta\cos\theta)\omega_{p(S)}-ck_{31(2)}$.\\
$E_{p(S)}$ & Lab & Slowly varying envelope of electric field of \\
            &    & pump (Stokes) laser. \\
$I_{p(S)}$ & Lab & Peak intensity of pump (Stokes) laser pulse \\
& & $I_{p(S)}=\frac{1}{2} c\epsilon_{0} E^{2}_{p(S)}$. \\
$I^{\mathrm{eff}}_{p(S)}$ & Lab & Effective peak intensity of pump (Stokes) \\
& & laser pulse, $I^{\mathrm{eff}}_{p(S)}=$\\
                          &   &$I_{p(S)}\frac{\Gamma}{\gamma(1+\beta\cos\theta)\Gamma_{p(S)}}$ \\
& & for $\Gamma<\gamma(1+\beta\cos\theta)\Gamma_{p(S)}$; \\
                          &   &$I_{p(S)}$ for $\Gamma\geq\gamma(1+\beta\cos\theta)\Gamma_{p(S)}$. \\
$T_{p(S)}$ & Lab & Pulse duration of pump (Stokes) laser.\\
$I_{1(2)}$ & Any & Angular momentum of ground state $\vert 1 \rangle$ ($\vert 2 \rangle$).\\
$L_{i3}  $ & Any & Multipolarity of the corresponding nuclear \\
& & $\vert i\rangle\rightarrow\vert 3\rangle$ transition.\\ 
$\mathbb{B}(\varepsilon/\mu\, L_{i3})$ & Rest & Reduced transition probability for the \\ 
 & & nuclear electric ($\varepsilon$) or  magnetic ($\mu$) \\
& &  $\vert i\rangle\rightarrow\vert 3\rangle$ transition.\\
%$\wp_{3i}$ & Dipole moment of $\vert3\rangle\rightarrow\vert i\rangle$ transition,\\
%           & $\wp_{3i}=\sqrt{\frac{3\epsilon_{0} hc^{3}B_{3i}\Gamma}{2W^{3}_{3i}}}$.\\
\hline
\hline
\end{tabular}
\end{table}
%\end{widetext}

%%%%%%%%%%%%%%%%%%%%%%%%%%%%%%%%%%%%%%%%%
\subsection{STIRAP versus $\pi$ pulses}
%%%%%%%%%%%%%%%%%%%%%%%%%%%%%%%%%%%%%%%%%

The interaction of a $\Lambda$-level scheme with the pump laser $P$ driving the $\vert 1\rangle\rightarrow\vert 3\rangle$  transition and the Stokes laser $S$ driving the $\vert 2\rangle\rightarrow\vert 3\rangle$  transition is depicted in Fig.~\ref{NCPT_threelevels}.  In STIRAP, at first the Stokes laser  creates a superposition of the two  unpopulated states $\vert 2 \rangle$ and $\vert 3 \rangle$.  Subsequently, the pump laser couples the fully occupied 
$\vert 1 \rangle$ and the pre-built coherence of the two empty states. If the two fields are sufficiently slowly varying and fulfill the adiabaticity condition, the dark (trapped) state 
\begin{equation}\label{darkstate}
\vert D\rangle = \frac{\Omega_{S}(t)}{\sqrt{\Omega_{p}^{2}(t)+\Omega_{S}^{2}(t)}}\vert 1\rangle-\frac{\Omega_{p}(t)}{\sqrt{\Omega_{p}^{2}(t)+\Omega_{S}^{2}(t)}}\vert 2\rangle. 
\end{equation}
is formed and evolves with the time-dependent pump and Stokes Rabi frequencies
$\Omega_{p}(t)$ and $\Omega_{S}(t)$, respectively \cite{bergmann1998}. Obviously, one can control the populations in the states $\vert 1\rangle$ and $\vert 2\rangle$ via temporally adjusting the laser parameters, e.g., the laser electric field strengths of pump and Stokes.
The adiabaticity condition for STIRAP states that 
\begin{equation}\label{ac}
\sqrt{\Omega_{p}^{2}+\Omega_{S}^{2}}\Delta\tau >10,
\end{equation}
where $\Delta\tau$ is the period during which the pulses overlap, and the  value of 10 on the right-hand side is rather empirical, based on numerical studies and experiments \cite{bergmann1998}.

Another option for achieving the coherent population transfer is utilizing so-called $\pi$-pulses. Using the scheme in Fig.~\ref{NCPT_threelevels}, let us consider the interaction of a two-level system with a single-mode laser, driving the $\vert 1\rangle\rightarrow\vert 3\rangle$ transition by the pump laser (for now we temporarily neglect state $\vert 2\rangle$ and the Stokes laser). The resulting state of this system is \cite{scully1997}
\begin{eqnarray}\label{pipulse}
\vert \psi\rangle&=&\cos\left( \frac{1}{2}\int_{-\infty}^{t}\Omega_{p}\left( \tau\right) d\tau\right) \vert 1\rangle
\nonumber \\
&+&\sin\left( \frac{1}{2}\int_{-\infty}^{t}\Omega_{p}\left( \tau\right) d\tau\right)\vert 3\rangle. 
\end{eqnarray}
Obviously, the complete coherent population transfer happens when
\begin{equation}
\int_{-\infty}^{t}\Omega_{p}\left( \tau\right) d\tau=n \pi
\end{equation}
for n odd. Because of this particular case, $\Omega_{p}\left( \tau\right)$ is called a $\pi$-pulse if $\int_{-\infty}^{\infty}\Omega_{p}\left( \tau\right) d\tau=\pi$. In the scheme in Fig.~\ref{NCPT_threelevels}, one can shine a pump $\pi$-pulse and subsequently a Stokes $\pi$-pulse on the target to coherently channel all population from state $\vert 1\rangle$ to state $\vert 2\rangle$ via the intermediate state $\vert 3\rangle$. This technique is termed as two $\pi$-pulses method.
One question may arise: why not directly pump the population from $\vert 1\rangle$ to state $\vert 2\rangle$ by using just one laser pulse? The advantages of the considered two-field scheme are \cite{bergmann1998}
that the
excitation efficiency can be made relatively insensitive to
many of the experimental details of the pulses. In addition,
with the three-state system, one can produce excitation
between states of the same parity, for which
single-photon transitions are forbidden for electric dipole
radiation, or between magnetic sublevels. Similarly in the case of nuclear coherent population transfer illustrated in Fig.~\ref{NCPT_threelevels}, either the direct transition between the two ground states is forbidden (e.g., the isomer state), or the required laser intensity for using one field will be higher than that of using two lasers due to the fact that the linewidth of the nuclear transition $\vert 1\rangle \rightarrow \vert 2\rangle$  may be much narrower than that of transition $\vert 1\rangle \rightarrow \vert 3\rangle$ and $\vert 2\rangle \rightarrow \vert 3\rangle$.

\begin{figure}
\begin{center}
\includegraphics[width=8.3cm]{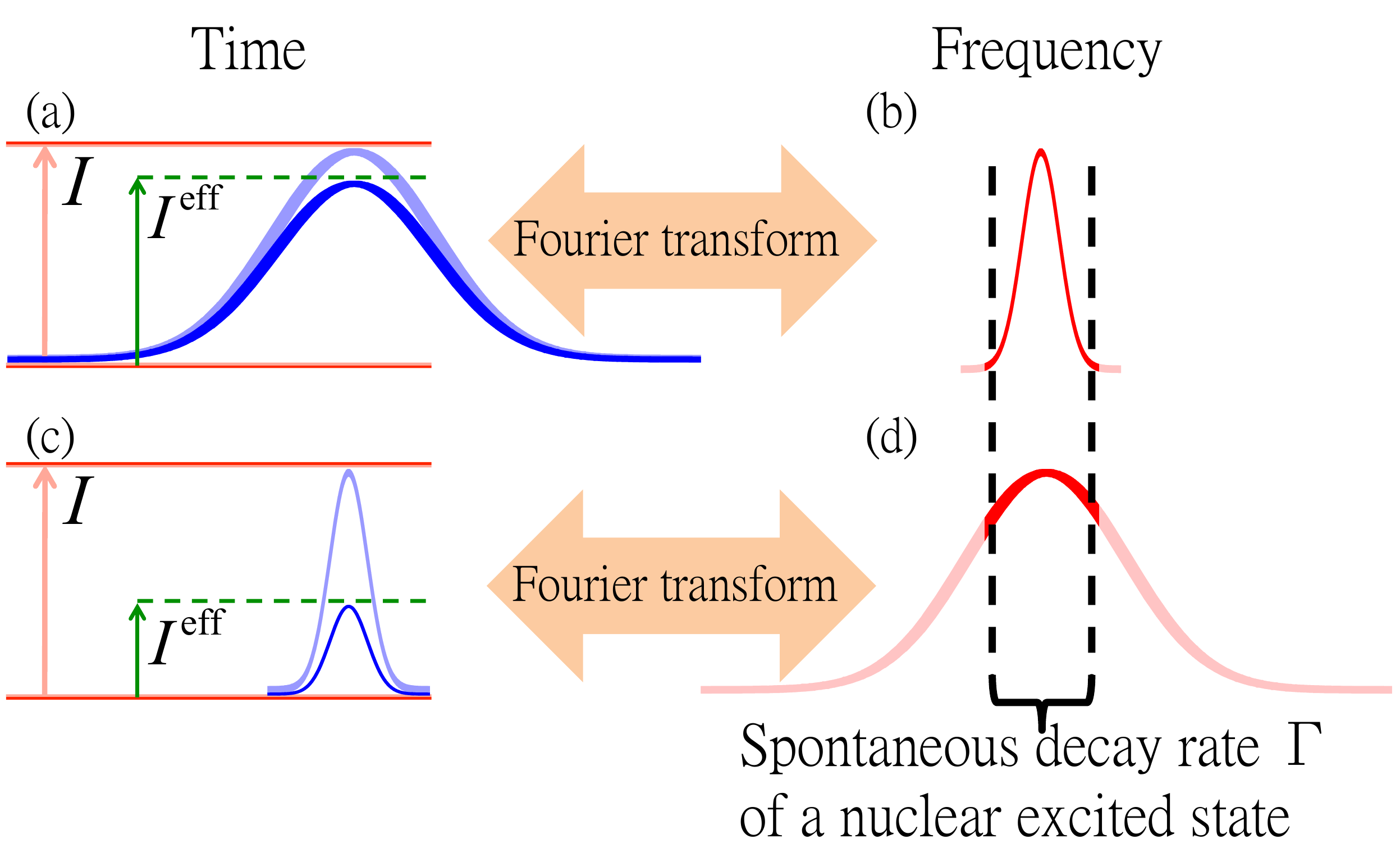}
\caption{The intuitive sketch of the concept of the effective intensity $I^{\mathrm{eff}}$. For the long laser pulse case in (a)(b), the bandwidth of the incident laser of intensity $I$ is narrower than the linewidth $\Gamma$ of the envisaged nuclear transition. For the short pulse case in (c)(d), the effective intensity is significantly reduced since the bandwidth of the incident laser is wider than $\Gamma$.} \label{NCPT_Ieff}
\end{center}
\end{figure}

The nuclear transition linewidth is typically narrower than the laser bandwidth. This property limits the resonant photon number within a produced XFEL pulse and gives a lower effective intensity \cite{palffy2008}. An intuitive picture of this issue is presented in Fig.~\ref{NCPT_Ieff}. The intensity $I$ of an incident XFEL pulse will not be fully observed by the nuclei, and the effective intensity $I^{\mathrm{eff}}$ depends on the ratio of the laser bandwidth to the corresponding nuclear transition linewidth. For instance, the effective intensity of a short incident pulse illustrated in Fig.~\ref{NCPT_Ieff}~(c)(d) is much weaker than that of a long pulse case (Fig.~\ref{NCPT_Ieff}~(a)(b)), because $\Gamma$ is wider than the bandwidth of the latter radiation pulse. 

\subsection{Nuclear decoherence caused by a partially coherent XFEL}
%%%%%%%%%%%%%%%%%%%%%%%%%%%%%%%%%%%%%%%%%%%%%%%%%%%%%%%%%%%%%%%%%%

\begin{figure}
\begin{center}
\includegraphics[width=8cm]{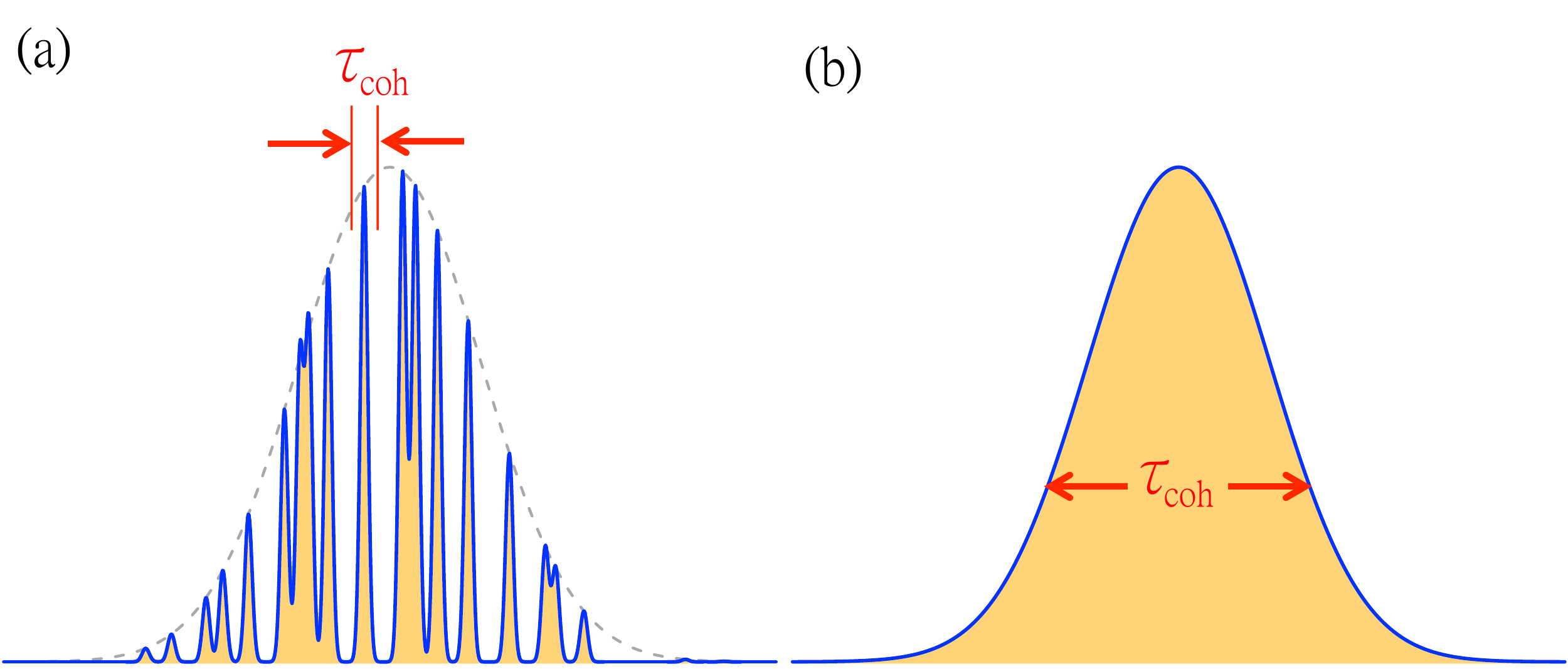}
\caption{Temporal profile of XFEL pulses with coherence time $\mathrm{\tau_{coh}}$. The yellow filled areas depict the temporal coherent part of a (a) partially, (b) fully coherent XFEL pulse. Additionally, the $\mathrm{\tau_{coh}}$ is shorter than the pulse duration of the gray dashed envelope in (a), whereas $\mathrm{\tau_{coh}}$ equals the pulse duration in (b). 
} \label{NCPT_XFELpulse}
\end{center}
\end{figure}

As we have seen in the discussion of STIRAP and $\pi$-pulses, the success of quantum coherent control using photons highly relies on the coherence of the laser beam. The role of coherence is obvious when noticing that both Eq.~(\ref{darkstate}) and Eq.~(\ref{pipulse}) depend on Rabi frequencies, which are proportional to the laser electric field, rather than the laser intensity \cite{bergmann1998}. This indicates that the whole nuclear dynamics will be driven by not only the number of incoming photons but also the phase of the laser beams. For discussing the role of the laser coherence time for the nuclear dynamics, we demonstrate two cases of XFEL pulses in Fig.~\ref{NCPT_XFELpulse}. In Fig.~\ref{NCPT_XFELpulse} (a), the temporal profile of a partially coherent XFEL reveals several coherent spikes \cite{altarelli2009}. Nuclear dynamics under the action of partially coherent pulses will experience several breaks or jumps between individual laser spikes. To describe such a quantum evolution, an additional dephasing term $\widehat{\rho}_{d}$  will be used to model the effects under the condition that the laser coherence time $\mathrm{\tau_{coh}}$ is shorter than the pulse duration. Then Eq.~(\ref{NCPT_eq1}) becomes \cite{buervenich2006}
\begin{equation}\label{NCPT_eqdephasing}
\frac{\partial}{\partial t}\widehat{\rho} = \frac{1}{i\hbar}\left[ \widehat{H},\widehat{\rho}\right]+\widehat{\rho}_{s}+\widehat{\rho}_{d}, 
\end{equation}
where the dephasing matrix
\begin{eqnarray}\label{dp}
\widehat{\rho}_{d} &=& -\frac{\gamma}{\mathrm{\tau_{coh}}}
\left( 
\begin{array}{ccc}
  0 & \eta_{D}\rho_{12} & \eta_{p}\rho_{13}\\
  \eta_{D}\rho_{21} & 0 & \eta_{S}\rho_{23}\\ 
  \eta_{p}\rho_{31} & \eta_{S}\rho_{32} & 0
\end{array}  
\right),
\end{eqnarray}
where $\eta_{D}=2+\beta+\beta\cos\theta_{S}$, $\eta_{p}=1+\beta$ and $\eta_{S}=1+\beta\cos\theta_{S}$.

To compensate the influence of a shorter $\mathrm{\tau_{coh}}$, a higher laser intensity is required. However, the above dephasing effect can be avoided by using a fully coherent XFEL with a pulse shape shown in Fig.~\ref{NCPT_XFELpulse} (b) to drive nuclear transitions. Since no phase jump occurs during the whole XFEL pulse duration,  the term $\widehat{\rho}_{d}$ in Eq.~(\ref{NCPT_eq1}) is absent.
%Both the success of STIRAP and that of two $\pi$-pulse are based on the full coherence of the laser pulses, therefore a fully coherent XFEL source is also paramount for nuclear coherent population transfer. 
Throughout the present study we assume a fully coherent XFEL source such as the future XFEL Oscillator (XFELO) \cite{kim2008} or the seeded XFEL (SXFEL) \cite{feldhaus1997,saldin2001,arthur2002,altarelli2009,yabashi2010,lcls2} for both pump and Stokes lasers. The complete coherence of the two laser beams ensures also their mutual coherence which is paramount for the creation of the dark state and the success of STIRAP.
%In the following, the fully coherent XFEL pulses are used together with acceleration of the target nuclei to achieve the resonance condition. The needed optimal laser intensity will be discussed for achieving not only 100\% population transfer but also coherent isomer triggering.
 
%%%%%%%%%%%%%%%%%%%%%%%%%%%%%%%%%%%%%%%%%%%%%%%%%%%%%%%%%%%%%%%%%%%%%%%%%
\section{Three-beam Setup\label{Setup}} 

\begin{figure}
\begin{center}
\includegraphics[width=7.5cm]{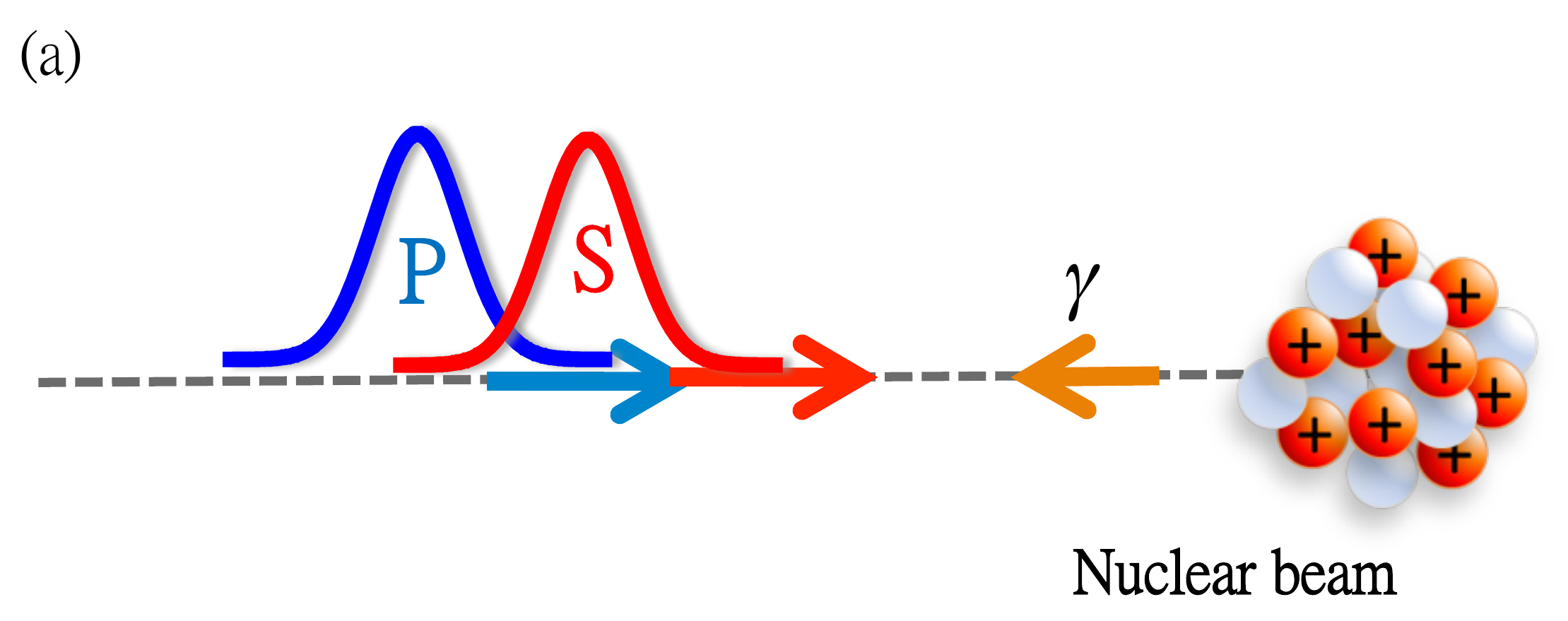}\hskip 0.1cm
\includegraphics[width=7.5cm]{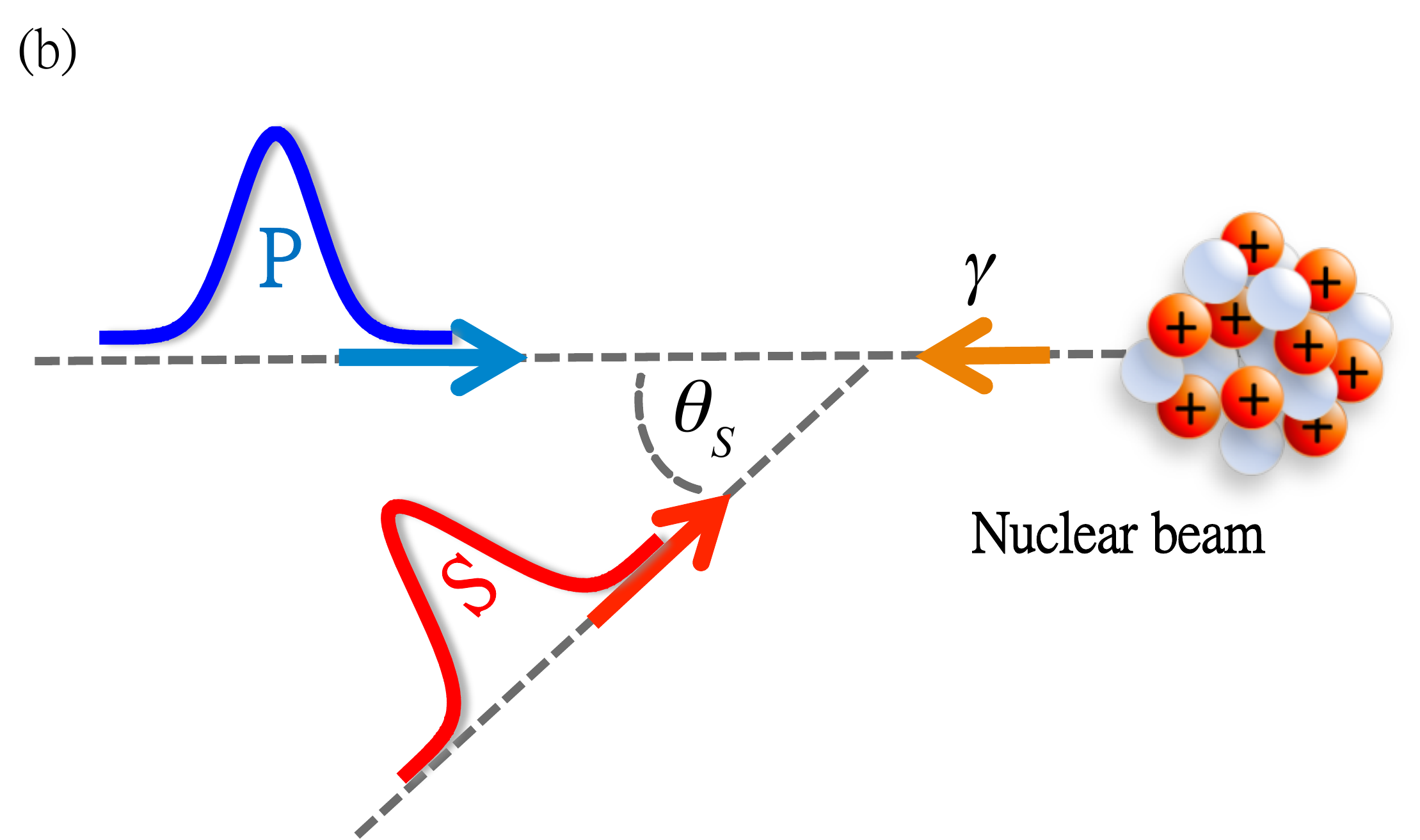}
\caption{
(a) Two-color scheme (copropagating-beams) in the lab frame. In this setup, the frequency of pump laser is different from that of the Stokes laser. The nuclear beam is accelerated such that $\gamma(1+\beta)\omega_{p(S)}=ck_{31(2)}$ is fulfilled. 
(b) One-color scheme (crossed-beams) in the lab frame. In this setup $\omega_{p}=\omega_{S}$, and the nuclear beam is accelerated such that both conditions $\gamma(1+\beta)\omega_{p}=ck_{31}$ and $\gamma(1+\beta \cos\theta_{S})\omega_{S}=ck_{32}$ are fulfilled. } \label{NCPT_setups}
\end{center}
\end{figure}

We study the collider system depicted in Fig.~\ref{NCPT_setups}, composed of an accelerated  nuclear beam that interacts with two incoming XFEL pulses.
The nuclear excitation energies are typically higher than the designed photon energy of the XFELO and SXFEL. The accelerated nuclei can interact with two Doppler-shifted x-ray laser pulses as showed in Fig.~\ref{NCPT_setups}~(a). The two laser frequencies (two-color) and the relativistic factor $\gamma$ of the accelerated nuclei have to be chosen such that in the nuclear rest frame both one-photon resonances are fulfilled. Copropagating laser pulses should have different frequencies in the laboratory frame in order to match the nuclear transition energies. To fulfill the resonance conditions with a one-color laser we envisage the pump and Stokes pulses meeting the nuclear beam at different angles ($\theta_S\ne 0$),  as shown in Fig.~\ref{NCPT_setups}~(b).

The most important prerequisite for nuclear STIRAP is the $temporal$ coherence of the x-ray lasers.  The coherence time of the existent XFEL at the Linac Coherent Light Source (LCLS) in Stanford, USA and of the European XFEL are on the order of 1~fs, much shorter than the pulse duration of 100~fs  \cite{emma2010,arthur2002,altarelli2009}. The SXFEL, considered as an upgrade for both facilities, will deliver completely transversely and temporally coherent pulses, that can reach 0.1~ps pulse duration and about 10 meV bandwidth \cite{saldin2001,lcls2}. Recently, a self-seeding scheme successfully produced a near Fourier-transform-limited x-ray pulses with $0.4-0.5$ eV bandwidth at $8-9$ keV photon energy \cite{amann2012}.
Another option is the XFELO that will provide  coherence time on the order of the pulse duration $\sim$~1~ps, and meV narrow bandwidth \cite{kim2008}.  We consider here the laser photon energy for the pump laser fixed at 25 keV for the XFELO and 12.4 keV for the SXFEL. 
The relativistic factor $\gamma$ is given by the one-photon resonance condition: 
\begin{equation}
E_{3}-E_{1}=\gamma(1+\beta)\hbar\omega_{p}. 
\end{equation}
Here we neglect the corrections due to recoil since they only introduce deviations on the order of $10^{-3}$ to the relativistic factor. 
The frequency of the Stokes x-ray laser  can be then determined depending on the geometry of the setup. For copropagating  pump and Stokes beams (implying a two-color XFEL), the photon energy of the Stokes laser is smaller than that of the pump laser since  $E_2>E_1$. The alternative that we put forward is to consider two crossed laser beams generated by a one-color SXFEL meeting the accelerated nuclei as shown schematically in Fig.~\ref{NCPT_setups}~(b). The angle  $\theta_{S}$ between the two beams is determined such that in the nuclear rest frame the pump and Stokes photons fulfill the resonances with the two different nuclear transitions. The values of 
$\gamma$, $\hbar\omega_S$ and $\theta_S$ for nuclear coherent population transfer for the nuclear systems under consideration are given in Table~\ref{NCPT_table3}. Here we extend the analysis in Ref.~\cite{liao2011} by investigating also the cases of nuclear coherent population transfer in $^{113}$Cd, $^{152}$Pm, $^{172}$Yb and $^{223}$Ra. 
The separation of the pump and Stokes beams out of the original XFEL beam requires dedicated x-ray optics such as the diamond mirrors  \cite{shvydko2010,shvydko2011,lindberg2011} developed for the XFELO. X-ray reflections can also help tune the intensity of the two beams. 
The coherence between the two ground states is crucial for successful nuclear coherent population transfer via STIRAP. Since in our case the lifetime of $\vert2\rangle$ is much longer than the laser pulse durations, decoherence is related to the unstable central frequencies and short coherence times of the pump and Stokes lasers.  Our one-color XFEL crossed-beam setup accommodates the present lack of two-color x-ray coherent sources and reduces the effect of laser central frequency jumps to equal detunings in the pump and Stokes pulses. However, the crossed-beam setup suffers from a number of drawbacks related to spatial and temporal overlap of the laser and ion beams, as it will be addressed shortly in the following.

%%%%%%%%%%%%%%%%%%%%%%%%%%%%%%%%%%%%%%%%%%%%%%%%%%%%%%%%%%%%%%%%%%%%%%%%%
\begin{widetext}
\begin{center}
\begin{table}[h]
\vspace{-0.4cm}
\caption{\label{NCPT_table3}
Nuclear, XFEL and nuclear beam parameters. $E_{i}$ is the energy of state $\vert i \rangle$ with $i\in \{1,2,3\}$ (in keV) \cite{nsdd}. The multipolarities and reduced matrix elements (in Weisskopf units) for the transitions $|j\rangle \rightarrow |3\rangle$ with  $j\in \{1,2\}$ are given. The accelerated nuclei have the relativistic factor $\gamma$, determined by the one-photon resonance condition $\gamma(1+\beta)\hbar\omega_{p}=ck_{31}$. For the copropagating-beams setup, $\hbar\omega_{S}$ denotes the Stokes photon energy.  The pump (copropagating beams) or both pump and Stokes lasers (crossed beams) photon energies are 12.4~keV for SXFEL and 25~keV for XFELO, respectively. For the crossed-beam setup, the angle $\theta_{S}$ between the pump and Stokes beams shown in Fig.~\ref{NCPT_setups}~(b) is given in rad.
}

\begin{tabular}{rrrrccccrrrrrr}
\hline
\hline
& & & &  \multicolumn{2}{c}{$\epsilon/\mu L$} &\multicolumn{2}{c}{$\mathbb{B}(\varepsilon/\mu\, L)$ (wsu)}& \multicolumn{3}{c}{SXFEL} & \multicolumn{3}{c}{XFELO} \\
\cline{5-14}
Nucleus&  $E_{3}$& $E_{2}$&$E_{1}$&  $| 1\rangle\rightarrow | 3\rangle$
&  $| 2\rangle\rightarrow |3\rangle$ &  $| 1\rangle\rightarrow | 3\rangle$  & $| 2\rangle\rightarrow |3\rangle$ & $\gamma$ & $\theta_{S}$ (rad) & $\hbar\omega_{S}$ (keV)&   $\gamma$ & $\theta_{S}$ (rad)& $\hbar\omega_{S}$ (keV)\\
\hline
$^{152}$Pm &44.45&25.02&0.00&E1&E1&$8.5\times 10^{-4}$&$2.8\times 10^{-4}$&1.9&1.7934&5.42&1.2&2.2740&10.93\\
$^{223}$Ra &50.13&29.86&0.00&E1&E1&$1.19\times 10^{-3}$&$5\times 10^{-4}$&2.1&1.8431&5.01&1.3&2.1965&10.11\\
$^{113}$Cd &522.26&316.21&263.54&E2&E1&$4.42\times 10$&$1.19\times 10^{-6}$&10.5&0.9373&9.88&5.2&0.9409&19.91\\
$^{185}$Re &284.00&125.00&0.00&E2&M1&$6.4\times 10$&$3.7\times 10^{-1}$&11.5&1.4544&6.93&5.7&1.4596&13.97\\
\cline{9-11}
$^{97}$Tc &657.00&324.00&96.57&E2&E1&$5\times 10^{2}$&$6.7\times 10^{-5}$&22.6&1.3836&7.36&11.2&1.3848&14.83\\
\cline{12-14}
$^{154}$Gd &1241.00&123.00&0.00&E1&E1&$4.4\times 10^{-2}$&$4.9\times 10^{-2}$&50.1&0.6407&11.17&24.8&0.6408&22.52\\
$^{172}$Yb &1599.87&78.74&0.00&E1&E1&$1.8\times 10^{-3}$&$1.23\times 10^{-3}$&64.5&0.4474&11.79&32.0&0.4475&23.77\\
$^{168}$Er &1786.00&79.00&0.00&E1&E1&$3.2\times 10^{-3}$&$9.1\times 10^{-3}$&72.0&0.4260&11.85&35.7&0.4260&23.88\\
\hline
\hline
\end{tabular}
\end{table}
\end{center}
\end{widetext}
%%%%%%%%%%%%%%%%%%%%%%%%%%%%%%%%%%%%%%%%%%%%%%%%%%%%%%%%%%%%%%%%%%%%%%%%%

The nuclear coherent population transfer via STIRAP is sensitive to the fulfillment of the two-photon resonance condition $\Delta_{p}=\Delta_{S}$. This involves on the one hand precise knowledge of the nuclear transition energy and on the other hand good control of  the laser frequency and therefore of the nuclear acceleration. The former is usually attained in nuclear forward scattering by scanning first for the position of the nuclear resonance. As for the latter, in our setup,  the relativistic factor $\gamma$ influences the detunings and the effective pump and Stokes intensities and Rabi frequencies.

So far  we have  considered an ideal case, using a monoenergetic beam from ion accelerators to bridge the gap between nuclear transition and x-ray laser energies. The designed $\gamma$ and $\theta_{S}$ are important parameters for achieving population transfer. However, for a realistic case both the beam divergency and the ion beam energy spread need to be considered, i.e., Eq.~(\ref{NCPT_eq1}) should be solved numerically with $\gamma\rightarrow\gamma+\Delta\gamma$ and $\theta_{s}\rightarrow\theta_{s}+\Delta\theta_{s}$. In the following, we attempt to connect the two-photon resonance condition $\Delta_{p}=\Delta_{S}$ to $\Delta\gamma$ and $\Delta\theta_{S}$ in the one-color setup illustrated in Fig.~\ref{NCPT_setups}~(b). Let us begin with the one photon detunings in the nuclear rest frame
\begin{eqnarray}
&&
\Delta_{p}=\omega_{p}\gamma(1+\beta)-ck_{31}\, ,
\nonumber\\
&&
\Delta_{S}=\omega_{S}\gamma(1+\beta\cos\theta_{S})-ck_{32}\, .
\label{NCPT_OneDetuning}
\end{eqnarray}
Substituting
\begin{eqnarray}
&&
\beta\rightarrow\sqrt{1-\frac{1}{\gamma^{2}}}\, ,
\nonumber\\
&&
\gamma\rightarrow\gamma+\Delta\gamma\, ,
\nonumber\\
&&
\theta_{S}\rightarrow\theta_{S}+\Delta\theta_{S}\, ,
\label{NCPT_Substituting}
\end{eqnarray}
into $\Delta_{p}=\Delta_{S}$,  we obtain
\begin{widetext}
\begin{eqnarray}
 \omega_{p}\left( \gamma+\Delta\gamma\right) \left[ 1+\sqrt{1-\frac{1}{\left( \gamma+\Delta\gamma\right) ^{2}}}\; \right]-ck_{31} =
\omega_{S}\left( \gamma+\Delta\gamma\right) \left[ 1+\sqrt{1-\frac{1}{\left( \gamma+\Delta\gamma\right)^{2}}}\cos\left( \theta_{S}+\Delta\theta_{S}\right) \right]-ck_{32}.
\end{eqnarray}
\end{widetext}
Using Taylor's expansion, and treating $\Delta\gamma$ and $\Delta\theta_{S}$ as small variables, the expansion becomes
\begin{widetext}
\begin{eqnarray}
& & \omega_{p}\gamma\left(1+\sqrt{1-\frac{1}{\gamma^{2}}}\;\right)+
\frac{\omega_{p}\Delta\gamma\left( 1+\sqrt{1-\frac{1}{\gamma^{2}}}\;\right)}{\sqrt{1-\frac{1}{\gamma^{2}}}}+\ldots-ck_{31} \nonumber\\
&=&\omega_{S}\gamma\left(1+\cos\theta_{S}\sqrt{1-\frac{1}{\gamma^{2}}}\;\right)-
\omega_{S}\gamma\Delta\theta_{S}\sin\theta_{S}\sqrt{1-\frac{1}{\gamma^{2}}}+\ldots
-ck_{32}.
\end{eqnarray}
\end{widetext}

Finally, we find the ideal 0\emph{$^{th}$} order two-photon resonance condition
\begin{equation}\label{NCPT_ZeroTwoPhoton}
\omega_{p}\gamma\left(1+\beta\right)-ck_{31}=\omega_{S}\gamma\left(1+\beta\cos\theta_{S}\right)-ck_{32},
\end{equation}
and the 1\emph{$^{st}$} order two-photon resonance condition (using $\omega_{S}=\omega_{p}$ for one-color setup)
\begin{equation}\label{NCPT_FirstTwoPhoton}
\Delta\theta_{S}=-\frac{1+\beta}{\beta^{2}\sin\theta_{S}}\left( \frac{\Delta\gamma}{\gamma}\right).
\end{equation}
Eq.~(\ref{NCPT_ZeroTwoPhoton}) is the condition for implementing STIRAP in an ideal case, i.e., the kinetic energy distribution of the nuclear beam is perfectly monoenergetic at the designed $\gamma$ ($\Delta\gamma=0$) and the divergence angle of the XFEL beam is zero ($\Delta\theta_{S}=0$). As mentioned before, in the real experiments the ideal condition is not fulfilled, e.g. the divergence angle of XFEL is on the order of 10$^{-6}$ rad \cite{altarelli2009}, and the velocity distribution of ion beams are not perfect. Then one has to consider the first order two-photon resonance Eq.~(\ref{NCPT_FirstTwoPhoton}) which gives a less strict STIRAP requirement, leading to an additional match between the non-monoenergetic nuclei and the photons that propagate along the direction at angles other than  $\theta_{S}$.

%%%%%%%%%%%%%%%%%%%%%%%%%%%%%%%%%%%%%%%%
\section{Numerical Results \label{NumRes}}
\begin{figure}
\begin{center}
\includegraphics[width=7cm]{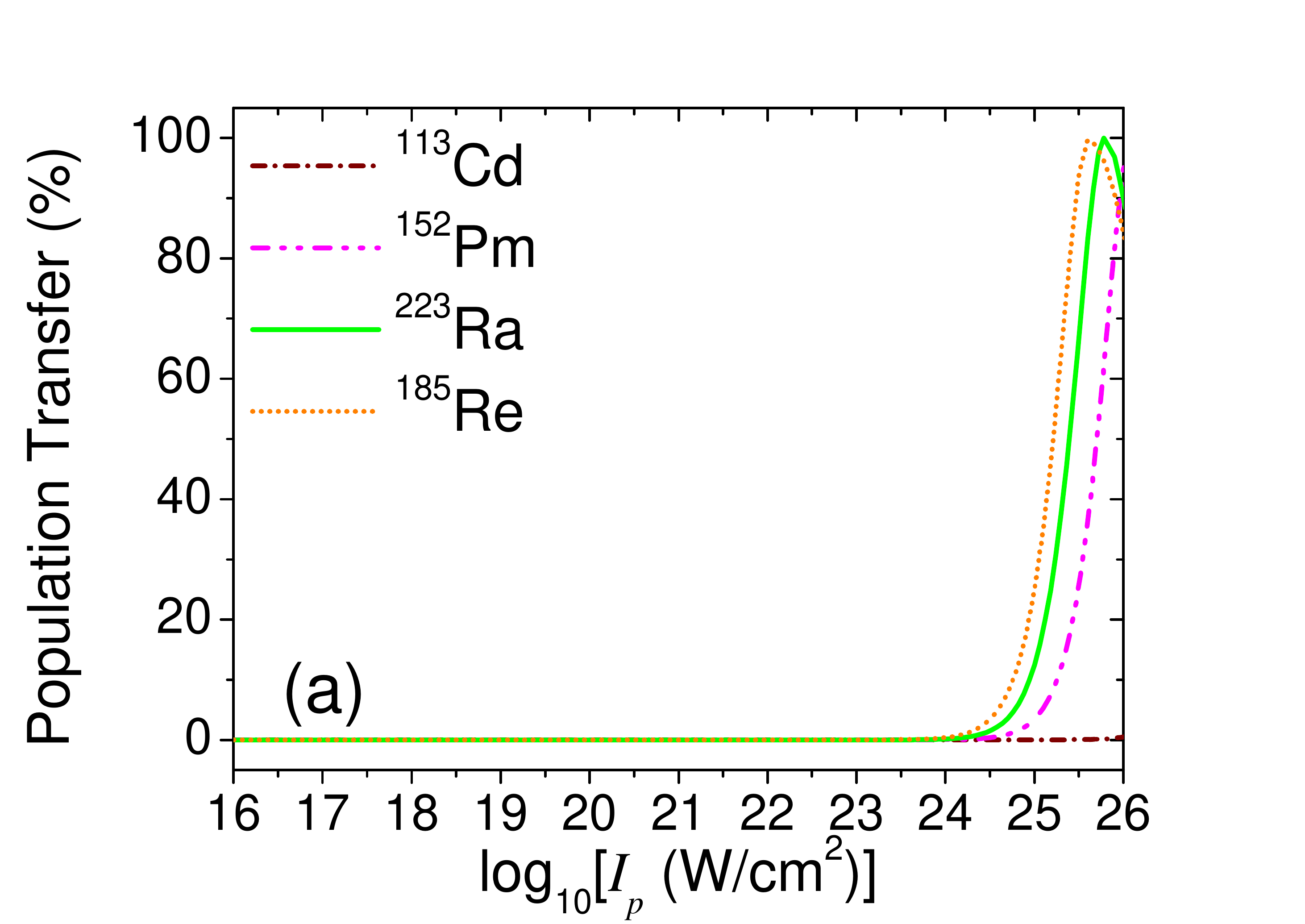}
\includegraphics[width=7cm]{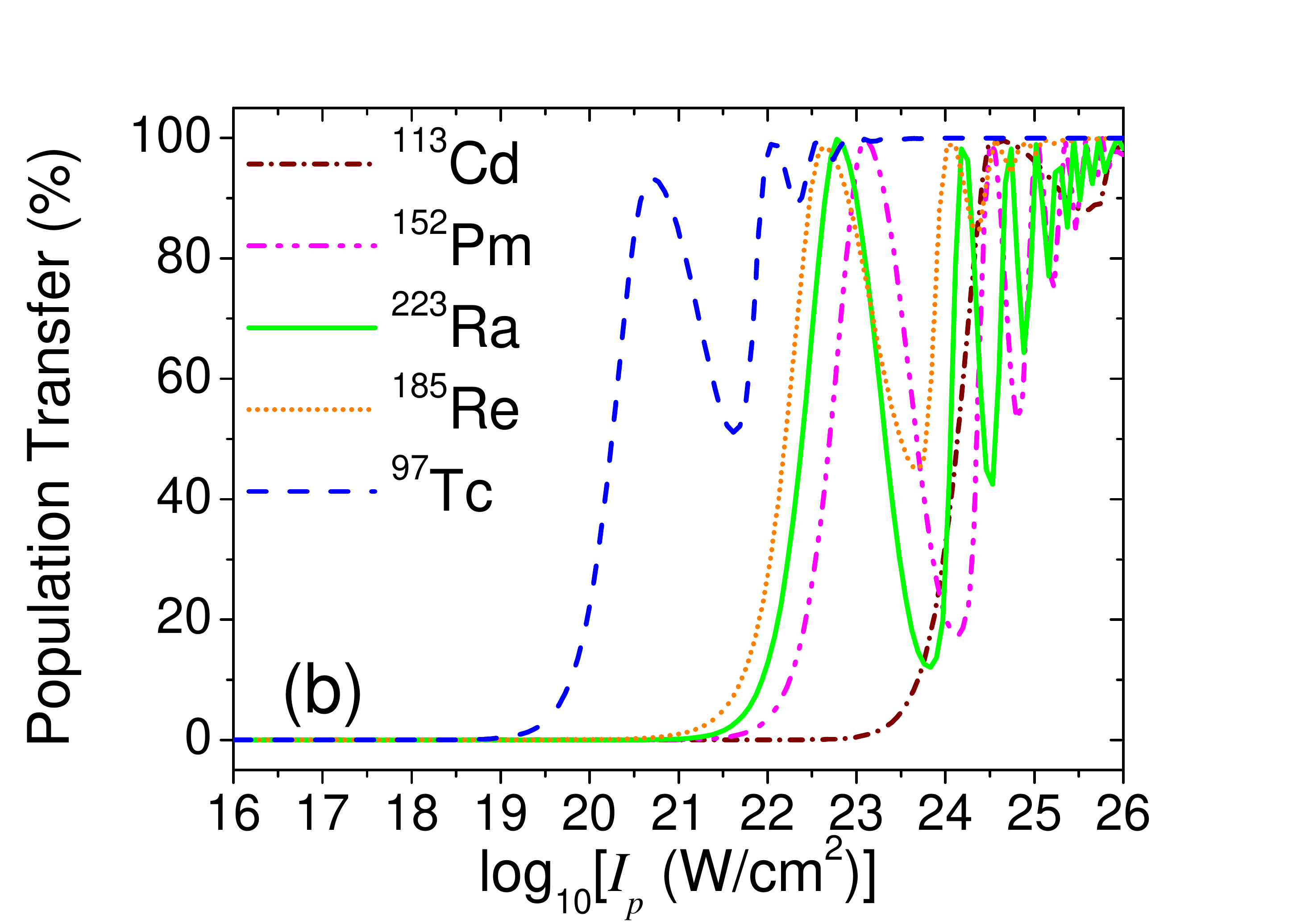}
\includegraphics[width=7cm]{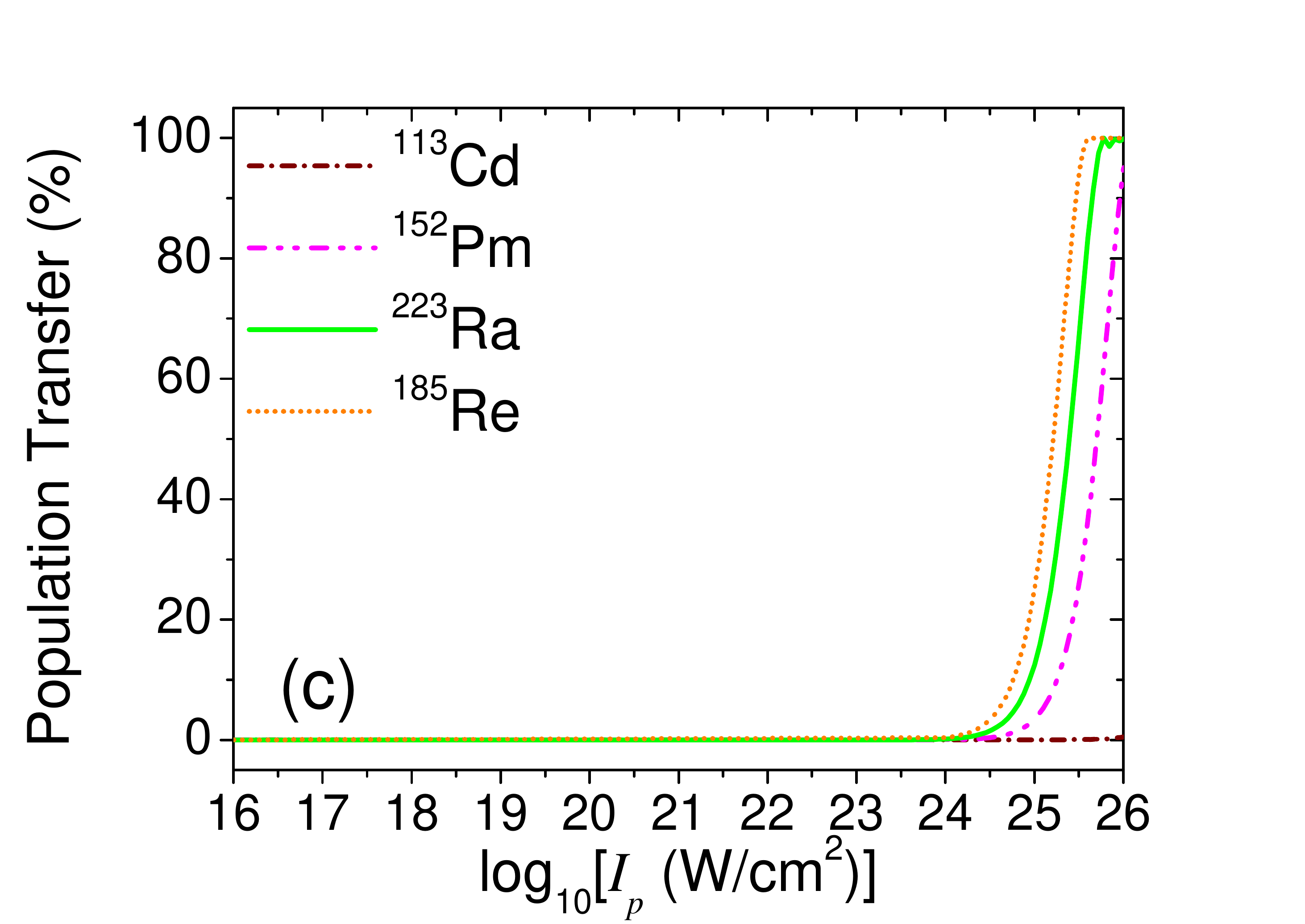}
\includegraphics[width=7cm]{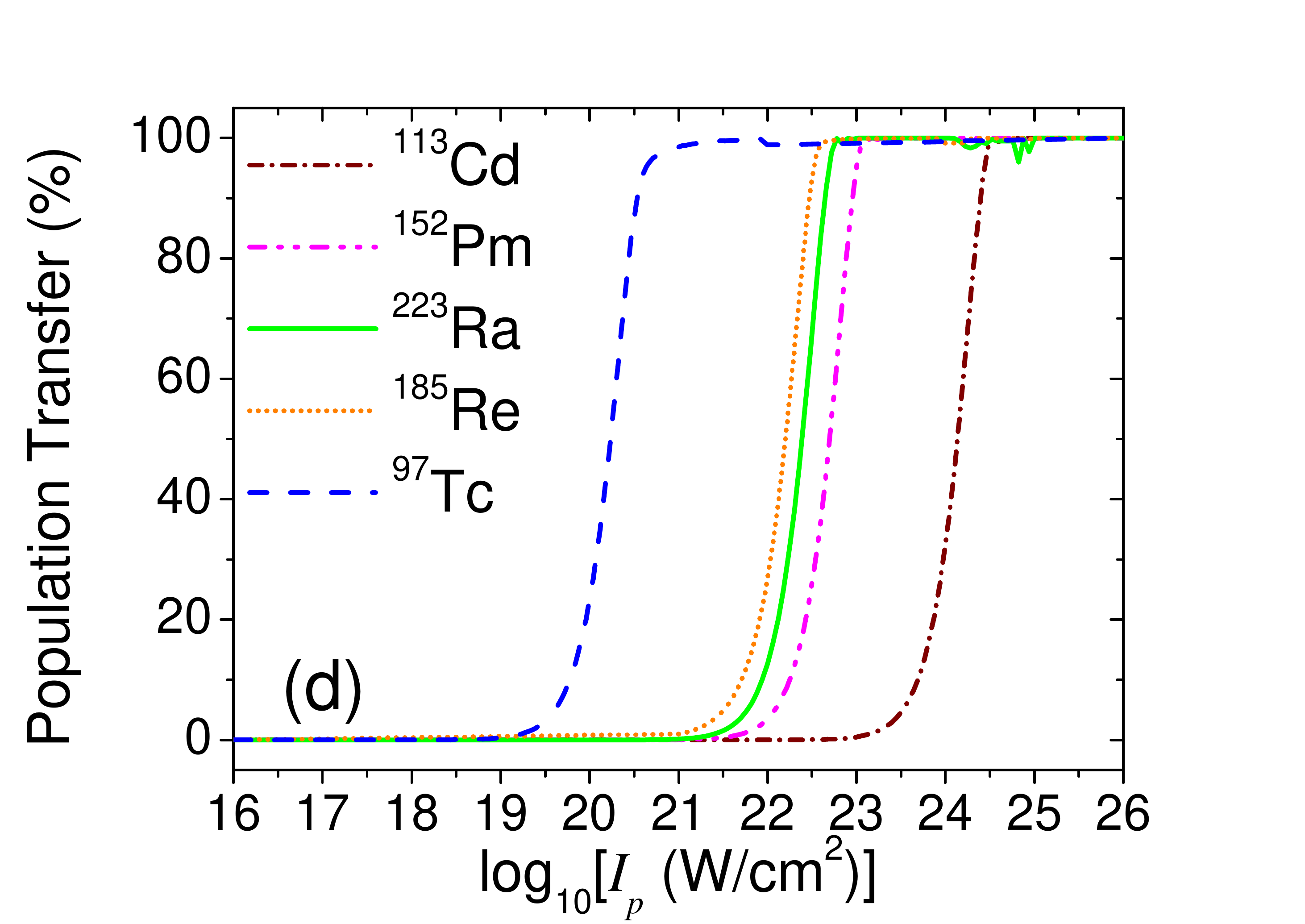}
\caption{Nuclear coherent population transfer for several nuclei with $\gamma<20$ as a function of the  pump XFEL peak intensity using  SXFEL (a,c) and  XFELO (b,d) parameters. For the crossed-beams setup (a) and (b), the Stokes laser intensities were chosen 
$I_{S}=32.81 I_{p}$ for $^{113}\mathrm{Cd}$,
$I_{S}=1.33 I_{p}$ for $^{152}\mathrm{Pm}$, 
$I_{S}=0.96 I_{p}$ for $^{223}\mathrm{Ra}$,
$I_{S}=0.02 I_{p}$ for $^{185}\mathrm{Re}$ and
$I_{S}=20.82 I_{p}$ for $^{97}\mathrm{Tc}$,
respectively, according to the $\pi$ pulse intensity ratios $I^{\pi}_{S}/I^{\pi}_{p}$. In the two-color setup (c) and (d), 
$I_{S}=41.19 I_{p}$ for $^{113}\mathrm{Cd}$, 
$I_{S}=3.04 I_{p}$ for $^{152}\mathrm{Pm}$, 
$I_{S}=2.38 I_{p}$ for $^{223}\mathrm{Ra}$, 
$I_{S}=0.03 I_{p}$ for $^{185}\mathrm{Re}$ and
$I_{S}=35.06 I_{p}$ for $^{97}\mathrm{Tc}$.  
All detunings are $\bigtriangleup_{p}=\bigtriangleup_{S}=0$. See discussion in the text and Tables~\ref{NCPT_table3} for further parameters.} \label{NCPT_intensitylowgamma}
\end{center}
\end{figure}

%%%%%%%%%%%%%%%%%%%%%%%%%%%%%%%%%%%%%%%%
In the following we present our nuclear coherent population transfer results for the nuclear three-level systems presented in Tables~\ref{NCPT_table3}.  We consider both the copropagating and crossed-beam setups presented in Fig.~\ref{NCPT_setups} and SXFEL and XFELO laser parameters. We differentiate between a higher $\gamma$ region ($\gamma\geq 20$) and a lower $\gamma$ region ($\gamma<20$) and present the percentage of nuclear population transfer as a function of the laser intensity in  Figs.~\ref{NCPT_intensityhighgamma}  and ~\ref{NCPT_intensitylowgamma}. The corresponding parameters $\gamma$, $\theta_{S}$ (for one-color setup), and Stokes photon energy $E_{S}$ (for two-color scheme) are  listed in Table~\ref{NCPT_table3}.

\subsection{High $\gamma$ Region ($\gamma\geq 20$)}
\begin{figure}
\begin{center}
\includegraphics[width=7cm]{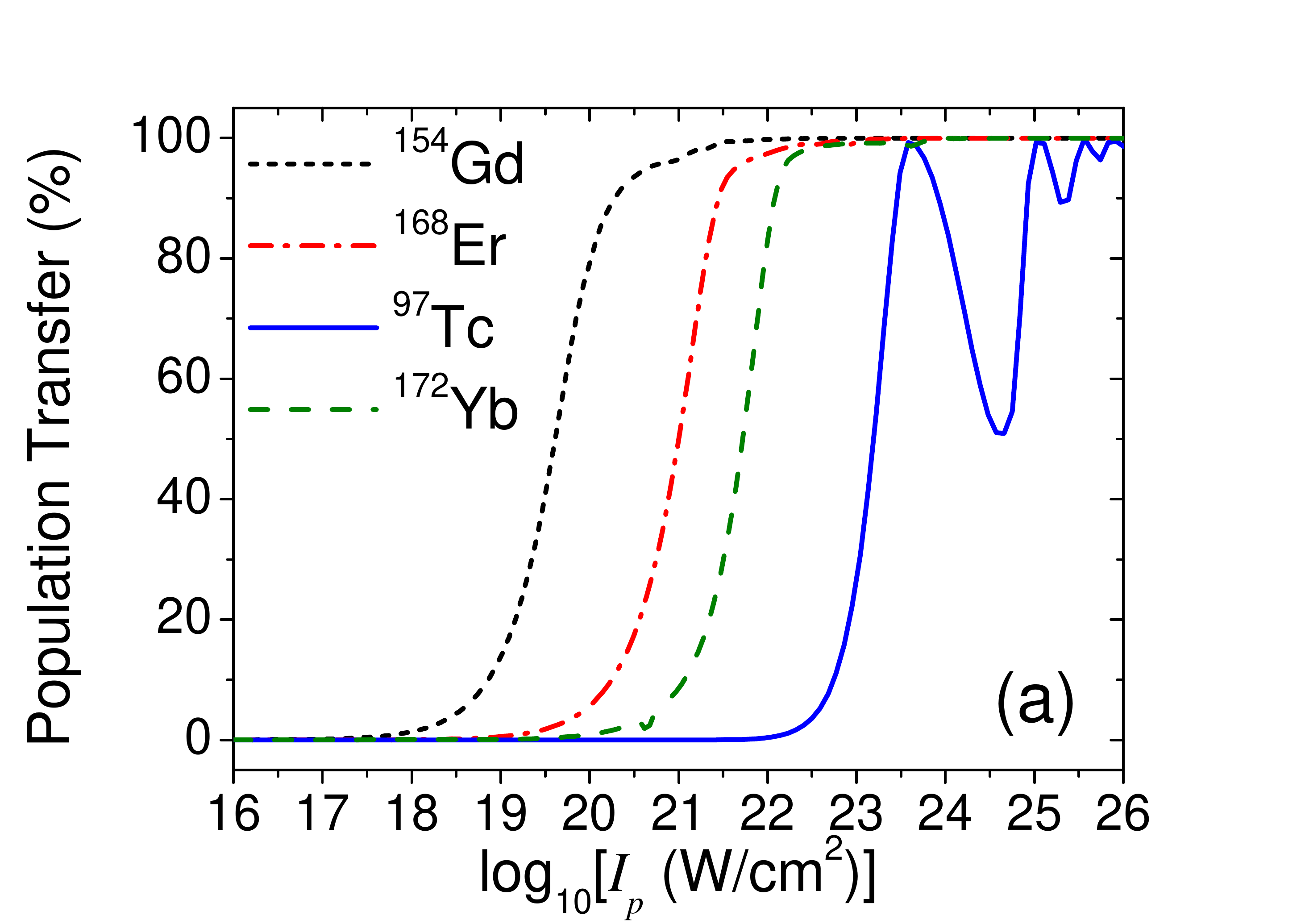}
\includegraphics[width=7cm]{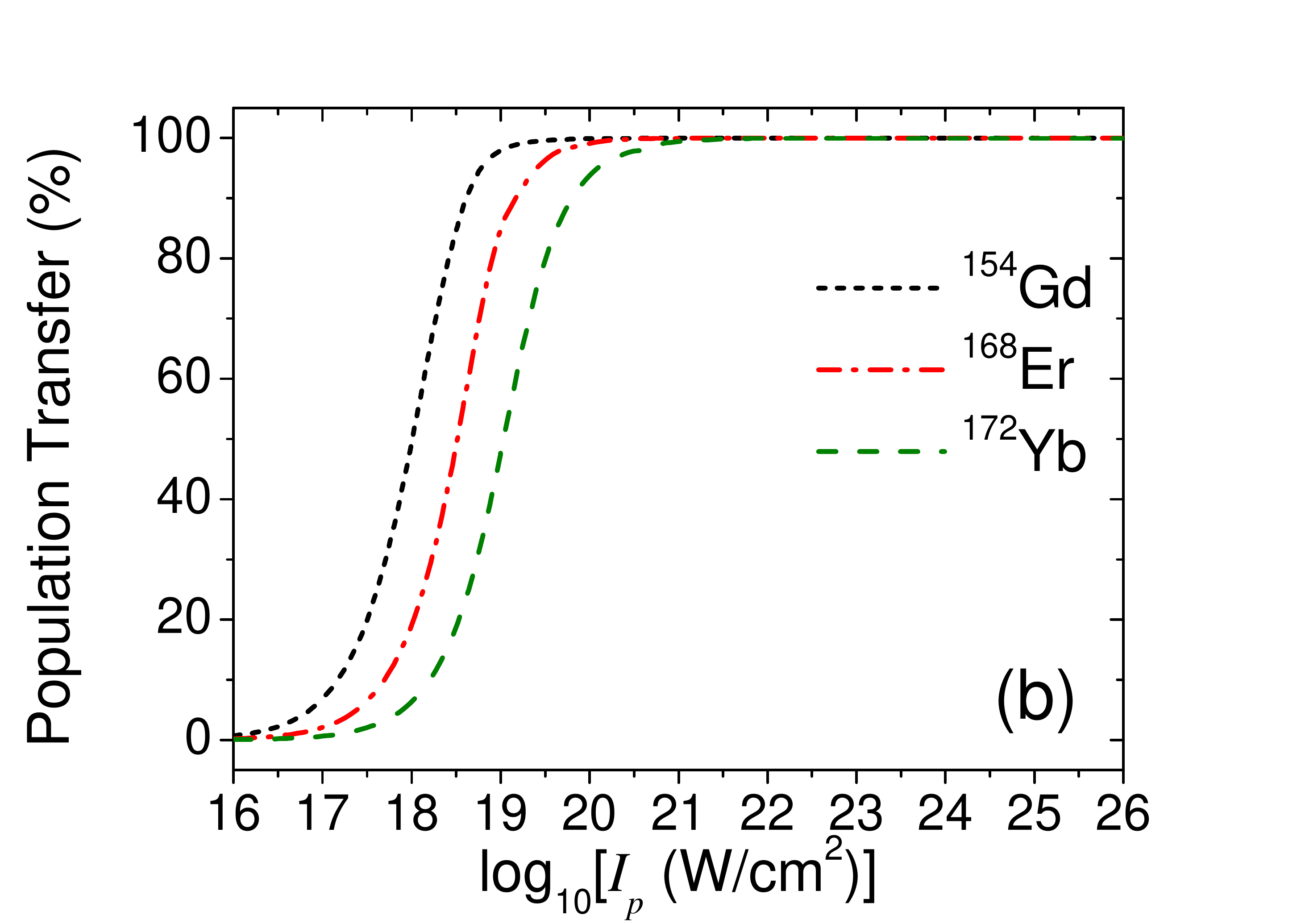}
\includegraphics[width=7cm]{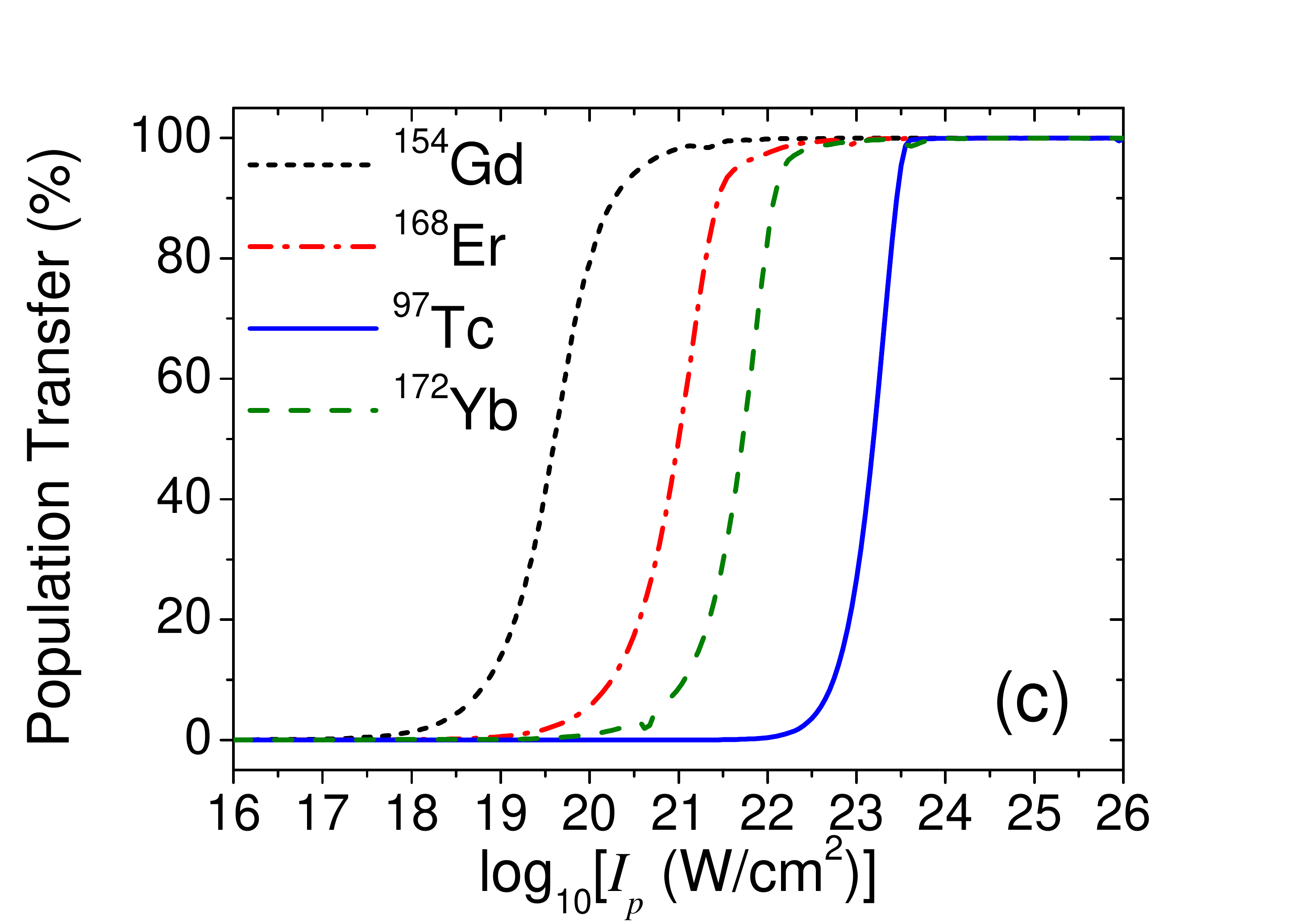}
\includegraphics[width=7cm]{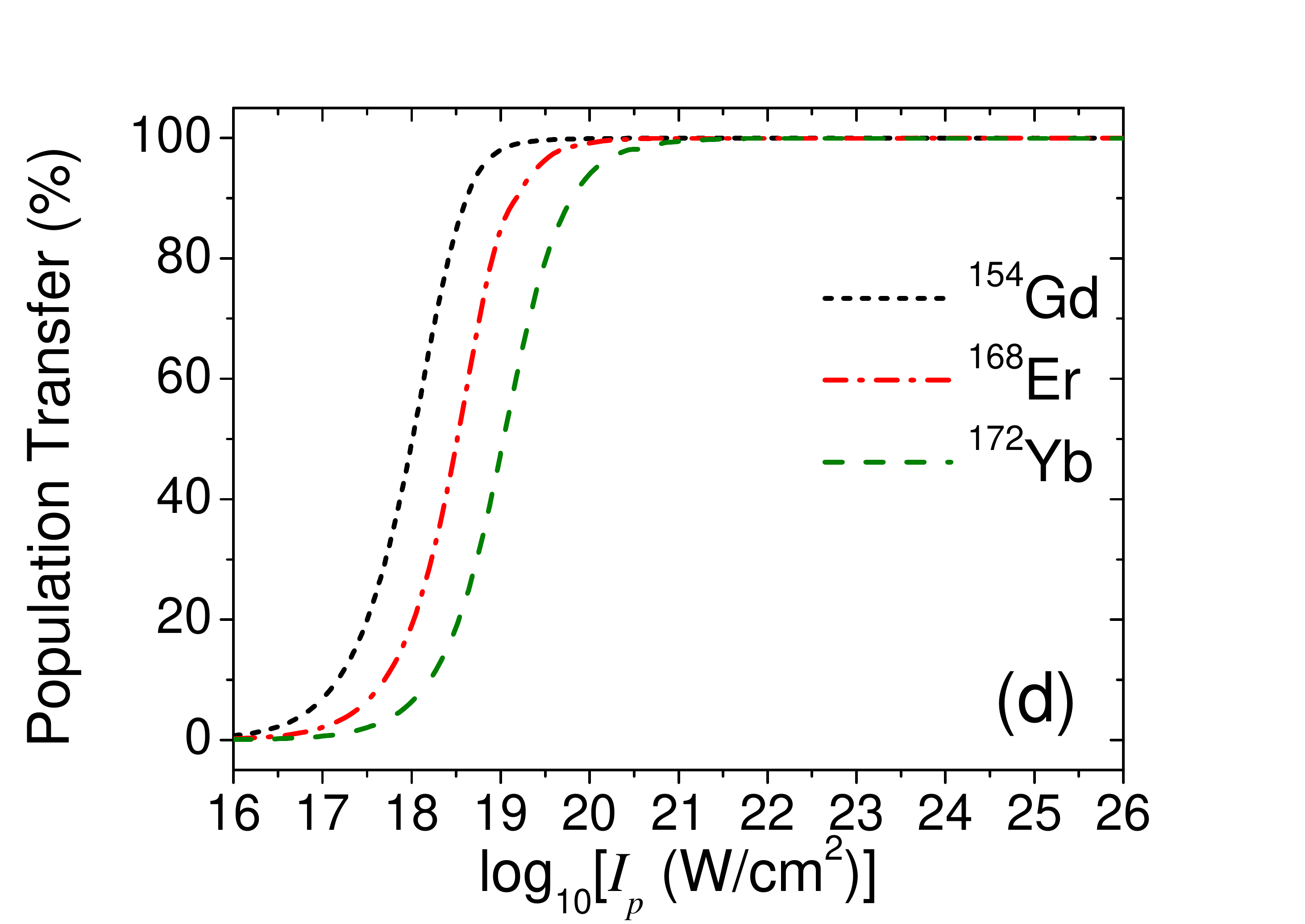}
\caption{Nuclear coherent population transfer for several nuclei with $\gamma\geq 20$ as a function of the  pump XFEL peak intensity using  SXFEL (a,c) and  XFELO (b,d) parameters. For the crossed-beams setup (a) and (b), the Stokes laser intensities were chosen 
$I_{S}=0.81 I_{p}$ for $^{154}\mathrm{Gd}$, 
$I_{S}=0.34 I_{p}$ for $^{168}\mathrm{Er}$,
$I_{S}=20.82 I_{p}$ for $^{97}\mathrm{Tc}$, 
$I_{S}=1.39 I_{p}$ for $^{172}\mathrm{Yb}$, and
respectively, according to the $\pi$ pulse intensity ratios $I^{\pi}_{S}/I^{\pi}_{p}$. In the two-color setup (c) and (d), 
$I_{S}=0.90 I_{p}$ for $^{154}\mathrm{Gd}$, 
$I_{S}=0.35 I_{p}$ for $^{168}\mathrm{Er}$, 
$I_{S}=35.06 I_{p}$ for $^{97}\mathrm{Tc}$ and
$I_{S}=1.46 I_{p}$ for $^{172}\mathrm{Yb}$. 
All detunings are $\bigtriangleup_{p}=\bigtriangleup_{S}=0$. See discussion in the text and Tables~\ref{NCPT_table3} for further parameters.} \label{NCPT_intensityhighgamma}
\end{center}
\end{figure}

\begin{figure}
\begin{center}
\includegraphics[width=8.3cm]{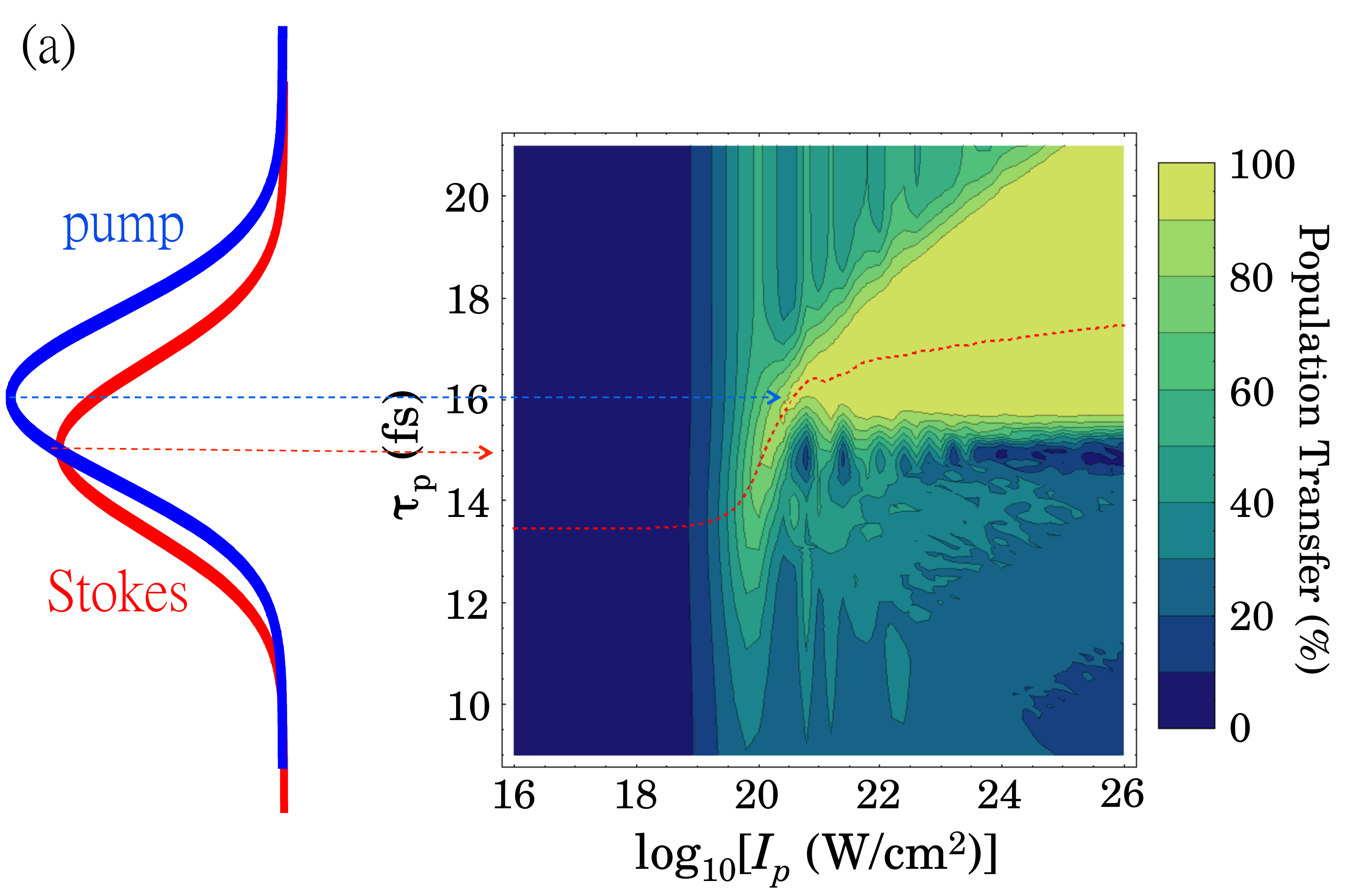}
\includegraphics[width=8.3cm]{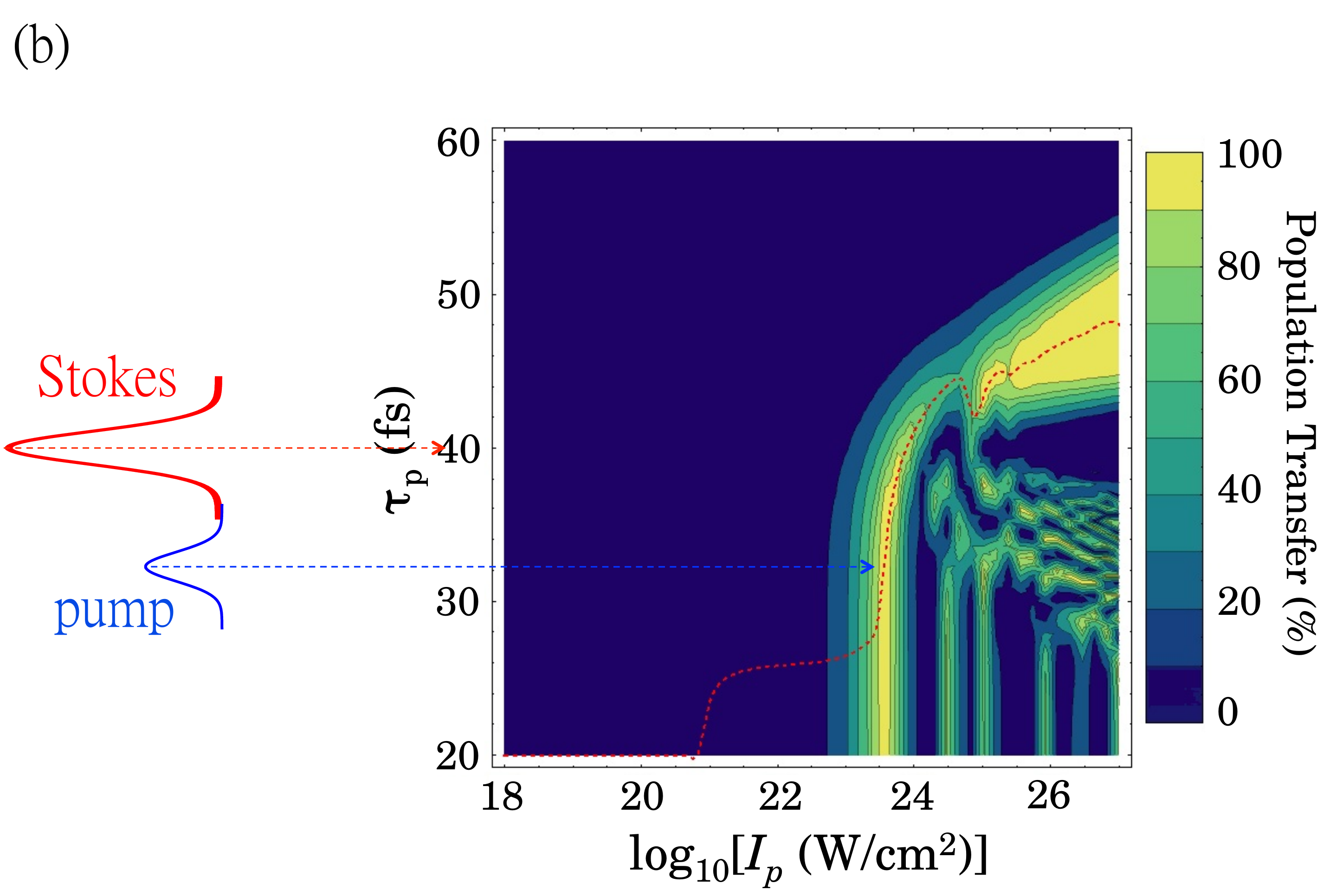}
\caption{The laser peak-intensity and pulse-delay dependent coherent population transfer. 
(a) for $^{154}$Gd, the Stokes peak position $\tau_{S}$ is fixed at 15 fs, and $I_{S}/I_{p}=0.81$. Eq.~(\ref{NCPT_eq1}) is numerically solved for $9\leq\tau_{p}\leq 21$ fs and $10^{16}\leq I_{p}\leq 10^{26}$ W/cm$^{2}$.
(b) for $^{97}$Tc, the Stokes peak position $\tau_{S}$ is fixed at 40 fs, and $I_{S}/I_{p}=20.82$. The numerical solution of Eq.~(\ref{NCPT_eq1}) is displayed in the interval of $20\leq\tau_{p}\leq 60$ fs and $10^{18}\leq I_{p}\leq 10^{27}$ W/cm$^{2}$. The nuclear population transfer curve of $^{154}$Gd and $^{97}$Tc in Fig.~\ref{NCPT_intensityhighgamma}~(a) are selected along the optimized red dashed lines in (a) and (b), respectively. 
} \label{NCPT_path}
\end{center}
\end{figure}

%%%%%%%%%%%%%%%%%%%%%%%%%%%%%%%%%%%%%%%%%%%%%%%%%%%%%
For this case, we present our results for $^{154}$Gd, $^{168}$Er, $^{97}$Tc (SXFEL parameters) and $^{172}$Yb that require stronger nuclear acceleration with $\gamma$ factors between 22 and 72 and fs pulse delays.   
For the one-color copropagating beams setup, the results  using SXFEL and  XFELO parameters are showed in Fig.~\ref{NCPT_intensityhighgamma}~(a) and Fig.~\ref{NCPT_intensityhighgamma}~(b), respectively.
The corresponding figures for the two-color crossed beam setup are Fig.~\ref{NCPT_intensityhighgamma}~(c) and Fig.~\ref{NCPT_intensityhighgamma}~(d).
The optimal set of laser parameters is obtained by a careful analysis of the dependence between pump peak intensity $I_{p}$ and pulse delay $\tau_{p}-\tau_{S}$.  A negative time delay corresponds to the $\pi$-pulse population transfer regime, while a positive one stands for STIRAP. For each value of $I_{p}$, the $\tau_{p}-\tau_{S}$ is chosen such that the nuclear population transfer reaches its maximum value. Two examples of this optimization process are demonstrated in Fig.~\ref{NCPT_path}.
Each red dashed line on the contour plot illustrates the optimal case, based on which we select the necessary SXFEL laser intensities showed in Fig.~\ref{NCPT_intensityhighgamma}~(a). For STIRAP, while the overlap time of the two pulses is very short, the effective Rabi frequencies $\sqrt{\Omega_{p}^{2}+\Omega_{S}^{2}}$ are in the considered examples very large, such that $\sqrt{\Omega_{p}^{2}+\Omega_{S}^{2}}\Delta\tau \geq 5$ already for 100\% nuclear coherent population transfer. 
Along the red dashed paths in Fig.~\ref{NCPT_path}, the adiabaticity condition (\ref{ac}) is fulfilled by slightly adjusting the laser intensity and the corresponding time delay between two XFEL pulses.

In Fig.~\ref{NCPT_path}, we clearly see  the working conditions for STIRAP and the two $\pi$-pulses method from the comparison between the lifetime of nuclear excited state $\vert 3\rangle$ and the laser pulse duration in the nuclear rest frame. For the case of $^{154}$Gd nucleus showed in Fig.~\ref{NCPT_path}~(a), the lifetime of  state $\vert 3\rangle$ is 1.54 fs \cite{nsdd} which is similar to the laser pulse duration of around 1 fs in the nuclear rest frame. Therefore the high coherent population transfer events for $^{154}$Gd only occur in the region of $\tau_{p}-\tau_{S}>0$, i.e., the STIRAP regime. For the opposite case when the lifetime of state  $\vert 3\rangle$ is  longer than the laser pulse duration in the nuclear rest frame, e.g., $^{97}$Tc (see Table~\ref{NCPT_table3} for the chosen levels) depicted in Fig.~\ref{NCPT_path}~(b), the high population transfer events can also happen in region $\tau_{p}-\tau_{S}<0$, i.e., the $\pi$-pulses regime.

The results for optimized intensity parameters in high $\gamma$ region are presented in Fig.~\ref{NCPT_intensityhighgamma}.
By using XFELO,
the $^{154}$Gd ground state population starts to be coherently transferred at  about $I_{p}=10^{17}$ W/cm$^{2}$, and alternatively at about $I_{p}=10^{19}$ W/cm$^{2}$ by using  SXFEL parameters. Up to $I_{p}=10^{19}$ W/cm$^{2}$ (XFELO) and $I_{p}=10^{21}$ W/cm$^{2}$ (SXFEL), more than 95$\%$ of the nuclei reach state $\vert 2 \rangle$. Due to the narrower bandwidth of XFELO, the required XFELO intensity for achieving 100\% population transfer is two orders of magnitude lower than that of SXFEL.
Additionally, in this case $\pi$ pulses cannot provide the desired nuclear coherent population transfer due to the fast spontaneous decay of state $|3\rangle$ in neither copropagating- nor cross-beam setups.
Similar behavior can be found in the cases of $^{168}$Er and $^{172}$Yb. These two species of nuclei are coherently channeled with XFELO intensity larger than $10^{17}$ W/cm$^{2}$, and reach 100\% population transfer at about $I_{p}=10^{20}$ W/cm$^{2}$. By utilizing SXFEL, all population of $^{168}$Er and $^{172}$Yb reach $|2\rangle$ at $I_{p}=10^{22}$ W/cm$^{2}$.
The calculated intensities necessary for complete nuclear coherent population transfer are within the designed intensities of the XFEL sources. Considering the operating and designed peak power of 20-100 GW \cite{emma2010,arthur2002,altarelli2009,saldin2001,lcls2} for SXFEL (and about three orders of magnitude less for XFELO) and the admirable focus achieved for x-rays of 7~nm \cite{mimura2009}, intensities could reach as high as 
 $10^{17}-10^{18}$ W/cm$^{2}$ for XFELO \cite{kim2008} and $10^{21}-10^{22}$ W/cm$^{2}$ for SXFEL \cite{saldin2001,lcls2}. As a recent development, focusing the 10 keV photon beam with the reflective optics at SACLA is expected to be applicable for the generation of a nm-size hard x-ray laser \cite{yumoto2012}. This progress may render possible an XFEL intensity larger than $10^{22}$ W/cm$^{2}$.

\subsection{Lower $\gamma$ Region ($\gamma<20$)}
For the lower $\gamma$ region, we demonstrate the population transfer results  for the cases of $^{113}$Cd, $^{152}$Pm, $^{223}$Ra, $^{185}$Re and $^{97}$Tc (XFELO parameters) that require less strong nuclear acceleration with $\gamma$ factors between 1 and 12. The results for optimized intensity parameters in lower $\gamma$ region are showed in Fig.~\ref{NCPT_intensitylowgamma}.
The smaller Doppler shift required in the lower $\gamma$ region corresponds to a smaller Lorentz boost of the laser electric field. As a consequence, 
the required laser intensity for achieving complete coherent population transfer is around three orders of magnitude higher than that in the high $\gamma$ region. (For investigating the pure nuclear response, in our calculation we have not included strong-field quantum electrodynamics effects such as pair creation \cite{piazza2012}.)
In the one-color setup, population transfer is achieved at lower intensities via sequential $\pi$ pulses. At the exact $\pi$-pulse value of the pump intensity, a peak in the nuclear population transfer for $^{185}$Re can be observed, at $I_p=6\times 10^{25}$ W/cm$^{2}$ in  Fig.~\ref{NCPT_intensitylowgamma}~(a) and $I_p=6\times 10^{22}$ W/cm$^{2}$ in  Fig.~\ref{NCPT_intensitylowgamma}~(b).   
With increasing $I_p$ in the crossed-beam setup (Fig.~\ref{NCPT_intensitylowgamma}~(a)(b)), the $^{185}$Re nuclei are only partially excited to state $|2\rangle$ and the population transfer yield starts to oscillate. The amplitude and frequency of the oscillations are varying as a result of our pulse delay optimization procedure.  At sufficient  intensities in the pulse overlap regime STIRAP becomes preferable as compared to the $\pi$ pulses mechanism due to the lack of oscillations. The plateau at 100$\%$ population transfer indicates that nuclear coherent population transfer via STIRAP alone is reached. 
A similar behavior is seen in the case of $^{152}$Pm and $^{223}$Ra. For  $^{113}$Cd, the required laser pulse intensity is about two orders of magnitude higher than  in the cases of the $^{152}$Pm, $^{223}$Ra and $^{185}$Re nuclei. This is due to the lower lower reduced transition rates $\mathbb{B}(E2)$ and $\mathbb{B}(E1)$ of $^{113}$Cd nuclei which result in a smaller Rabi frequency.
In the two-color copropagating beams scheme presented in Fig.~\ref{NCPT_intensitylowgamma}~(c)(d), the pulse shape of pump and Stokes pulse are the same in the nuclear rest frame. This makes STIRAP more efficient and thus preferable  compared to the single-color setup, as the STIRAP plateau can be reached  with lower laser intensities.

\subsection{Coherent Isomer Triggering}
%%%%%%%%%%%%%%%%%%%%%%%%%%%%%%%%%%%%%%%%
One of the most relevant applications of nuclear coherent population transfer is isomer triggering. In Fig.~\ref{NCPT_intensityhighgamma} and \ref{NCPT_intensitylowgamma} we present our results  for nuclear coherent population transfer in $^{97}$Tc nuclei starting from the $E_1=96.57$~keV isomeric state which has a half life of $\tau_1=91$~d. 
Due to the state $|3\rangle$ life time ($\geq 0.76$ ps \cite{nsdd}) of $^{97}$Tc nuclei is longer than the laser pulse duration 
such that population transfer at lower intensities can be achieved via $\pi$ pulses in the crossed-beam setup. The intensity for which complete isomer depletion is achieved  using SXFEL is $I_{p}=4\times10^{23}$ W/cm$^{2}$. Additionally, due to the longer pulse duration of the XFELO and consequently higher losses via spontaneous decay of state $|3\rangle$, the peak population transfer at $I_p=5.2\times 10^{20}$ W/cm$^{2}$ reaches only 93$\%$ in Fig.~\ref{NCPT_intensitylowgamma}~(b) in the crossed-beam setup. For the copropagating beams setup and XFELO parameters, 100$\%$ nuclear coherent population transfer is achieved for the same intensity  $I_p=5.2\times 10^{20}$ W/cm$^{2}$ as illustrated in Fig.~\ref{NCPT_intensitylowgamma}~(d).

Moreover, we investigate the coherent population transfer in $^{113}$Cd nuclei for isomer depletion. 
The nuclear population starts from the $E_1=263.54$~keV isomeric state which has a half life of $\tau_1=14.1$~y.
The numerical results are presented in Fig.~\ref{NCPT_intensitylowgamma}~(b)(d) with XFELO parameters (the required SXFEL intensity is too large to be illustrated in Fig.~\ref{NCPT_intensitylowgamma}~(a)(c)). Comparing with the case of $^{97}$Tc, the required laser intensity for complete coherent isomer triggering for $^{113}$Cd is around four orders of magnitude higher. This is caused by the lower reduced transition rates $\mathbb{B}(E2)$ and $\mathbb{B}(E1)$ of $^{113}$Cd nuclei together with the lower $\gamma$ factors.

\subsection{Realistic Beam Parameters}
\begin{figure}
\begin{center}
\includegraphics[width=6cm]{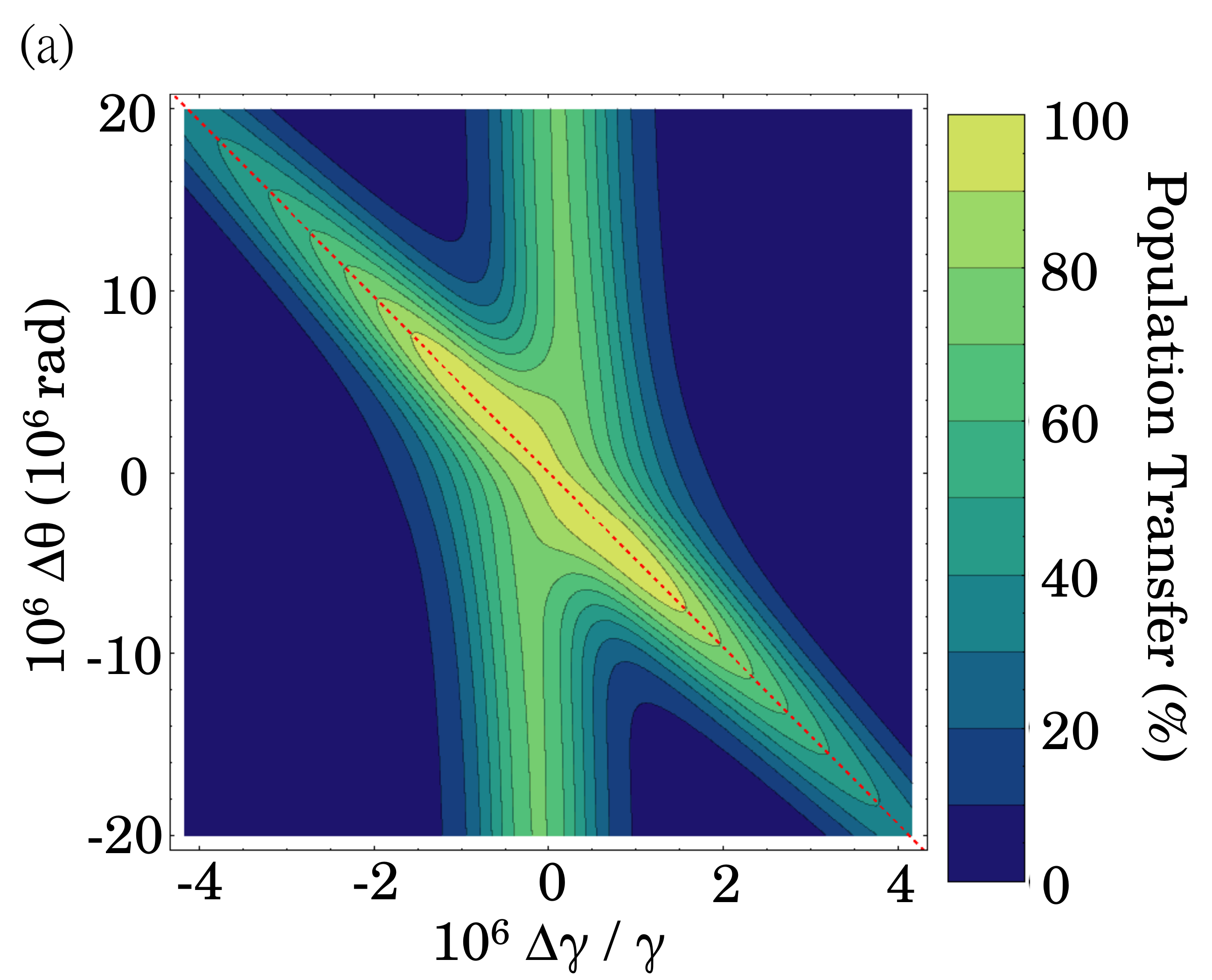}
\includegraphics[width=6cm]{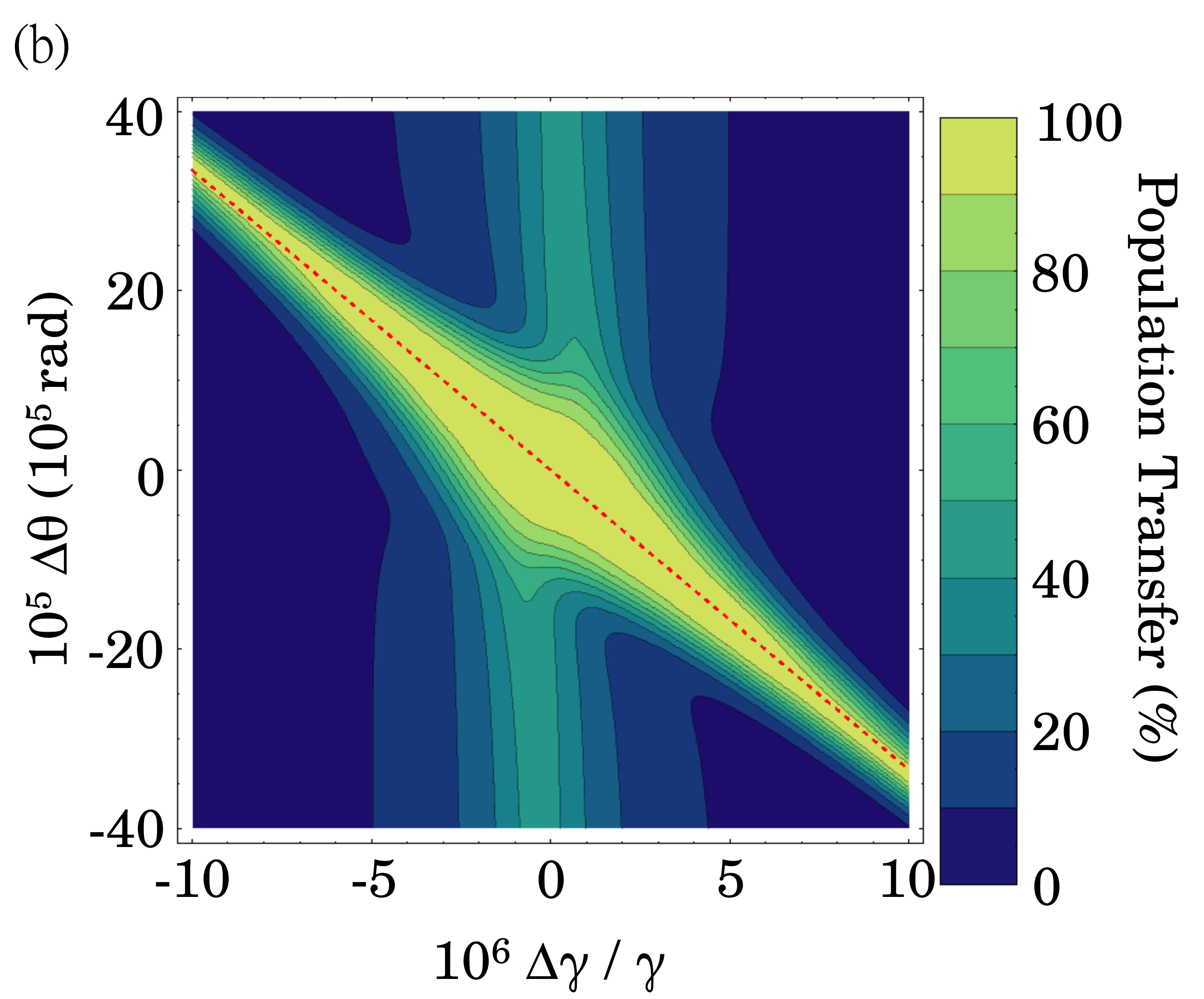}
\caption{$\Delta\theta_{S}$- and $\Delta\gamma$-dependent nuclear coherent population transfer in the one-color scheme. (a) For $^{168}$Er, $I_{p}=2.85\times 10^{22}$ W/cm$^{2}$ and $I_{S}=9.57\times 10^{21}$ W/cm$^{2}$. (b) For $^{154}$Gd, $I_{p}=6.78\times 10^{21}$ W/cm$^{2}$ and $I_{S}=5.49\times 10^{21}$ W/cm$^{2}$. The color coding shows the different percentage of population transfer with SXFEL. The red dashed line depicts Eq.~(\ref{NCPT_FirstTwoPhoton}).} \label{NCPT_errors}
\end{center}
\end{figure}

%%%%%%%%%%%%%%%%%%%%%%%%%%%%%%%%%%%%%
We now turn to the implementation of a realistic case including both beam divergence and ion velocity spread. 
In the low $\gamma$ region, the forthcoming FAIR at GSI will provide high quality ion beams with energies up to 45 GeV/u \cite{fair2006}. The corresponding $\gamma$ limit is about 48 and the precision $\Delta E/E\sim 2\times10^{-4}$. For the high $\gamma$ region, the Large Hadron Collider (LHC) is currently the only suitable ion accelerator which can accelerate $^{208}$Pb$^{82+}$ up to $\gamma=2963.5$ with low energy spread of about $10^{-4}$ \cite{lhc2011}. LHC  can also accelerate lighter ions to energies larger than 100 GeV \cite{carminati2004}. 
For the strong acceleration regime, the resonance condition corresponds to an energy spread of the ion beam of 
$10^{-5}$. This issue becomes more problematic for nuclei that require the moderate acceleration regime where the resonance condition requires a more precise $\gamma$ value, $\Delta \gamma/\gamma=10^{-6}$. 
On the other hand, the European XFEL will deliver laser pulses with the divergence angle of about $10^{-6}$ rad \cite{altarelli2009}.  This causes the missmatch of $\Delta_p\neq\Delta_S$ together with the energy spread $\Delta E$ of an ion beam. To address the realistic case in Fig.~\ref{NCPT_setups}~(b), we numerically solve Eq.~(\ref{NCPT_eq1}) with $\gamma\rightarrow\gamma+\Delta\gamma$ and $\theta_{s}\rightarrow\theta_{s}+\Delta\theta_{s}$.
Our results are presented in Fig.~\ref{NCPT_errors}, where the two errors $\Delta\gamma$ and $\Delta\theta_{s}$ are scanned for the cases of $^{154}$Gd and $^{168}$Er.
Eq.~(\ref{NCPT_FirstTwoPhoton}) is 
illustrated by the red dotted line in Fig.~\ref{NCPT_errors}, and the agreement is verified by comparing it with the high nuclear population transfer region of the numerical solution. 
We find the nuclear coherent population transfer maintains values of around $80\%$ in the region of $\theta_{S}\pm 10^{-5}$ rad and $\Delta \gamma/\gamma=\pm 10^{-6}$ for $^{154}$Gd and $^{168}$Er. 

%%%%%%%%%%%%%%%%%%%%%%%%%%%%%%%%%%%%%%%%%%%%%%%%%%%%%%%%%%%%%%%%%%%%%%%%%

\section{Experimental Facilities} \label{ExpFacil}

\subsection{Large Infrastructure and Table-Top Solutions}
X-ray coherent light sources, listed in Table~\ref{NCPT_table4}, are not available today at the few large ion acceleration facilities. At present  a new materials research center MaRIE (Matter-Radiation Interactions in Extreme) providing both a fully coherent XFEL with photon energy of 50 keV  and accelerated charged-particle beams is envisaged in the USA \cite{marie2011}. In addition, the photonuclear physics pillar of the Extreme Light Infrastructure (ELI) can provide simultaneously 
a compact XFEL as well as ion acceleration  reaching up to 4-5 GeV \cite{eli2011}. 
At ELI, the combination of gamma-rays and acceleration of the nuclear target are already under consideration 
for  nuclear resonance fluorescence experiments \cite{eli2011}. Furthermore, ELI is also envisaged to deliver gamma rays with energies of few MeV \cite{eli2011}, which could be used for direct photoexcitation of giant dipole resonances 
\cite{weidenmueller2011}.

Tabletop solutions for both ion acceleration and x-ray coherent light, showed in Table~\ref{NCPT_table5}, would facilitate the experimental realization of isomer depletion by nuclear coherent population transfer and nuclear batteries. Tabletop x-ray  undulator sources are already operational \cite{fuchs2009}, with a number of ideas envisaging compact x-ray  FELs \cite{nakajima2008,gruener2007}. Rapid progress   spanning five orders of magnitude increase in the achieved light brightness within only two years has been reported \cite{kneip2008,kneip2010,kneip2012}.
In conjunction with the crystal cavities  designed for the XFELO, such table-top devices have the potential to become a key tool for the release on demand of energy stored in nuclei at large ion accelerator facilities. Alternatively, the exciting forecast of compact shaped-foil-target ion
accelerators \cite{chenb2009,zhang2012b}, foil-and-gas target \cite{zheng2012} and radiation pressure  acceleration \cite{hegelich2002,willingale2006,bulanov2010,carroll2010,ter2012} together with microlens beam focusing \cite{toncian2006} are likely to provide a viable table-top solution to be used together with the existing large-scale XFELs.

%%%%%%%%%%%%%%%%%%%%%%%%%%%%%%%%%%%%%%%%%%%%%%%%%%%%%%%%%%%%%%%%%%%%%%%%%
\begin{widetext}
\center{
\begin{table}[h]
\vspace{-0.4cm}
\caption{\label{NCPT_table4}
Specifications of the existent and forthcoming XFEL facilities.
}

\center{
\begin{tabular}{l|ccccc}
\hline
\hline
Facility & LCLS  & SACLA & European XFEL & MaRIE & XFELO\\
\hline
Ref. & \cite{arthur2002,vartanyants2011,gutt2012}&\cite{yabashi2010,yumoto2012}&
\cite{altarelli2009}&\cite{marie2011}&\cite{kim2008}\\
\hline
Location & SLAC  & Spring-8 & DESY      & LANL  & ANL   \\
(Country)& (USA) & (Japan)  & (Germany) & (USA) & (USA) \\
\hline
Status & Operation & Operation & Construction & Proposal &Proposal\\
\hline
Photon      & $5-25$ & $<$15.5 & $0.25-12.4$ & 50 & $5-20$\\
energy (keV)&        &       &             &    &       \\
\hline
Coherence & $0.55-2$ & Not  & $0.2-1.4$ & Not  & 1000\\
time (fs) &          & clear&           & clear&     \\
\hline
Pulse         & $10-300$ & $<$100 & 100 & $<$100 & 1000\\
duration (fs) &          &        &     &        &     \\
\hline
Peak      & $\sim$10 & $5-29$ & $20-150$ & 10     & 0.001\\
power (GW)&          &        &          &        &      \\
\hline
Photon      & $10^{11}-10^{12}$ & $5\times 10^{11}$ & $10^{12}-10^{14}$ & $10^{11}$& $10^{9}$\\
per pulse   &                   &                   &                   &          &         \\
\hline
Beam size             & $1.3\times 1.3-$  & $0.95\times 1.2-$ & $55\times 55-$ & Not   & Not  \\
(FWHM) ($\mu$m$^{2}$) & $3000\times 3000$ & $33\times 33$     & $90\times 90$  & clear & clear\\
\hline
\hline
\end{tabular}
}

\end{table}}
%\end{widetext}
%%%%%%%%%%%%%%%%%%%%%%%%%%%%%%%%%%%%%%%%%%%%%%%%%%%%%%%%%%%%%%%%%%%%%%%%%

%%%%%%%%%%%%%%%%%%%%%%%%%%%%%%%%%%%%%%%%%%%%%%%%%%%%%%%%%%%%%%%%%%%%%%%%%
%\begin{widetext}
\center{
\begin{table}[h]
\vspace{-0.4cm}
\caption{\label{NCPT_table5}
Specifications of some table-top x-ray sources.
}

\center{
\begin{tabular}{l|ccccc}
\hline
\hline
%Scheme & Plasma wiggler & Thomson back scattering & Magnet undulator & HHG (AMO) & Carbon nanotube\\
Scheme & Plasma  & Thomson back & Magnet    & HHG  & Carbon\\
       & wiggler & scattering   & undulator & (AMO)& nanotube\\
\hline
Ref. & \cite{kneip2008,kneip2010,cipiccia2011,kneip2012}&\cite{schwoerer2006,karagodsky2010}
&\cite{gruener2007,nakajima2008,schlenvoigt2007,fuchs2009}&\cite{chen2010}&\cite{bagchi2011}\\
\hline
Photon      &$1-150$&$0.4-1000$&0.14&$>$0.5&$50-500$\\
energy (keV)&       &          &    &    &          \\
\hline
Coherent&Spatially&Not&Spatially&Spatially&Not\\
\hline
Peak      &$10^{21}-10^{23}$&$3\times 10^{17}$&$1.3\times 10^{17}$&$6\times 10^{7}$ photons/s&Not clear\\
brilliance&                 &                 &                   &                          &         \\
\hline
Pulse         &$10-30$&Not clear&10&0.01&Not clear\\
duration (fs) &       &         &  &    &         \\
\hline
\hline
\end{tabular}
}
\end{table}}
\end{widetext}
%%%%%%%%%%%%%%%%%%%%%%%%%%%%%%%%%%%%%%%%%%%%%%%%%%%%%%%%%%%%%%%%%%%%%%%%%

\begin{figure}[h]
\begin{center}
\includegraphics[width=8cm]{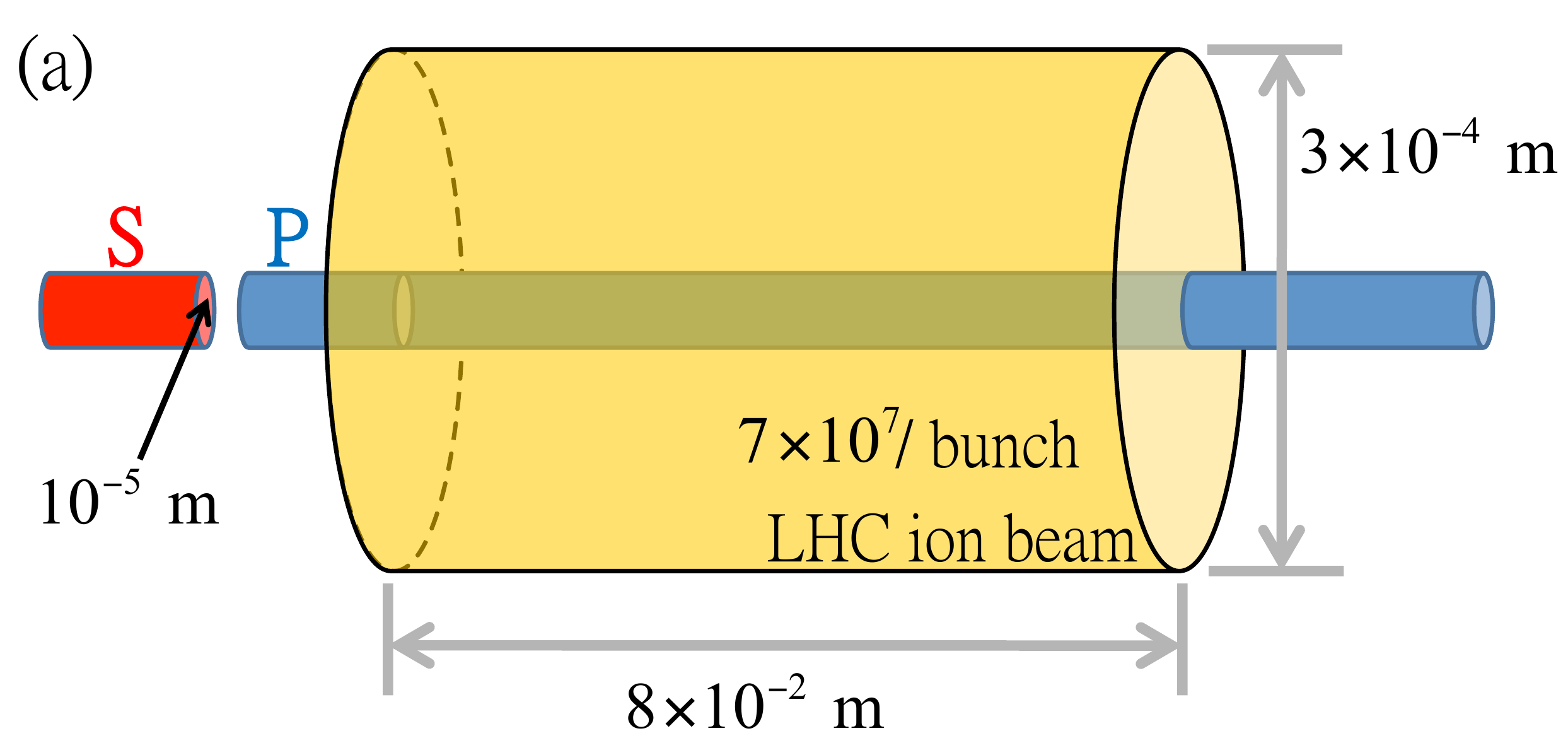}
\includegraphics[width=8cm]{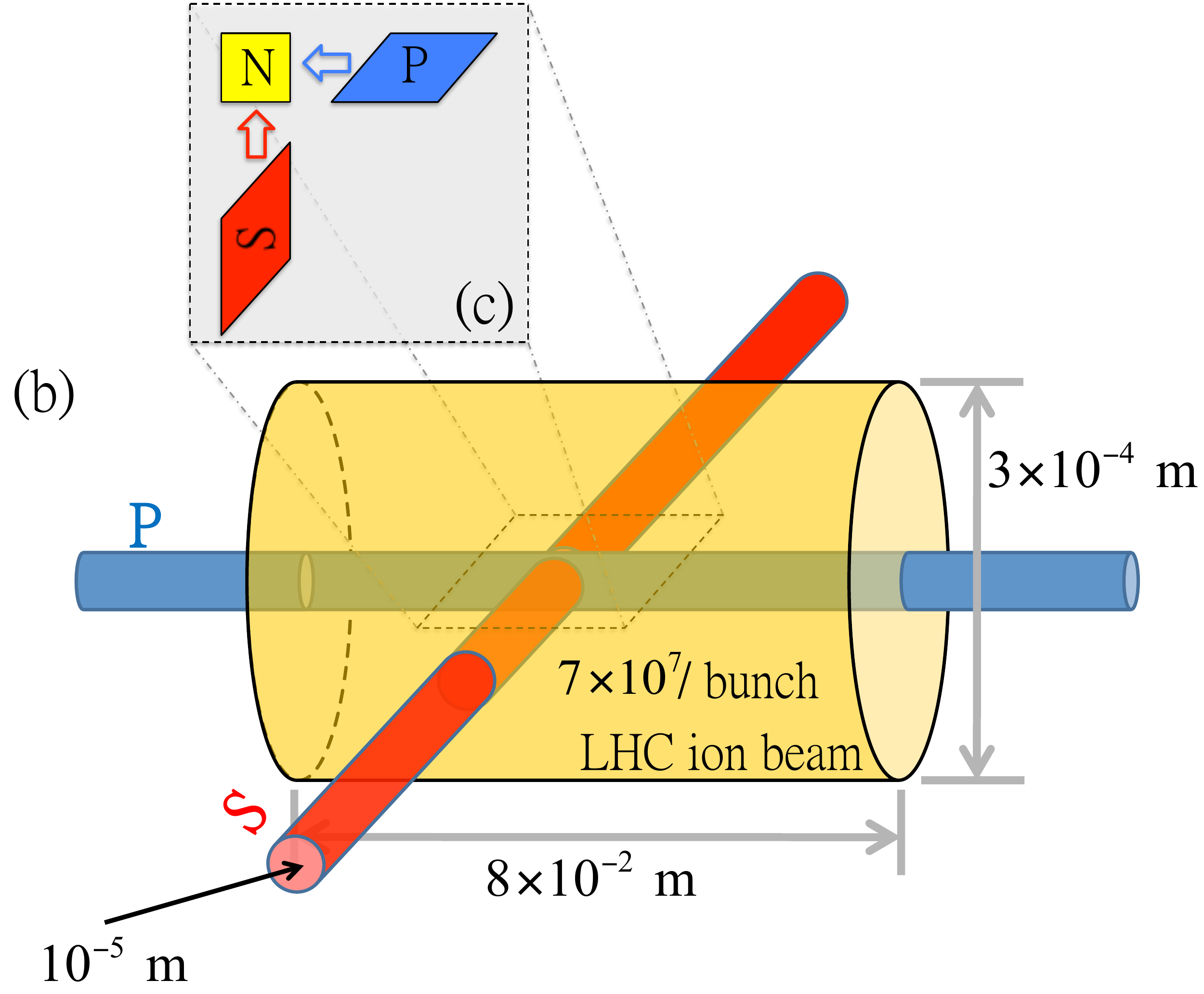}
\caption{ The spatial overlap between the nuclear beam, pump pulse and the Stokes pulse in the nuclear rest frame for (a) the two-color scheme and (b) the one-color scheme. The parameters of the LHC ion beam are considered \cite{lhc2011}. The blue (red) `pipeline' is the volume that the pump (Stokes) XFEL pulse flies through, and the light orange cylinder denotes the nuclear bunch. (c) the required XFEL wave front for implementing STIRAP in a one-color setup. The blue (red) parallelogram depicts the wave front of pump (Stokes) pulse and the yellow square illustrates the nuclear area.} \label{NCPT_SpatialOverlap}
\end{center}
\end{figure}
\subsection{The Spatial Overlap between the XFEL and Nuclear Beams}
A further study of the overlap efficiency for the laser beams and ion bunches shows that the copropagating laser beams setup is more advantageous. In Fig.~\ref{NCPT_SpatialOverlap}, using the LHC beam size parameters \cite{lhc2011} and a 10~$\mu$m focusing of the XFEL beam, we estimate that for (a) copropagating laser beams up to $10^5$ nuclei meet the laser focus per bunch and laser pulse, while for (b) crossed laser beams this number reduces to 30. The extreme temporal and spatial fine-tuning required to match the overlaps of a bunched ion beam with the two laser beams in the crossed-beam setup is however at present out of reach. A continuous ion beam, on the other hand, has the disadvantage of much lower ion density at the overlap with the pump and Stokes beam. Furthermore, the necessary time delay between pump and Stokes and the adiabaticity condition  for STIRAP will be in this case only fulfilled for ions at the diagonal line of the overlap area. In order to maintain the pulse delay and the 
adiabaticity condition for the whole overlap region with the nuclear beam, a special laser pulse front as presented in Fig.~\ref{NCPT_SpatialOverlap}~(c) is required. We conclude therefore that for a number of technical and conceptual reasons, the two-color copropagating beams scheme has better chances to be realized experimentally in the near future.

%%%%%%%%%%%%%%%%%%%%%%%%%%%%%%%%%%%%%%%%%%%%%%%%%%%%%%%%%%%%%%%%%%%%%%%%%

\section{Conclusion} \label{sec:NCPT-summary}
In summary, we have investigated nuclear coherent population transfer using a collider system composed of two fully coherent XFEL beams together with an ion accelerator, and considered the interaction between an accelerated nuclear bunch and two XFEL pulses. This system is showed to be  a powerful tool for studying the radiation-nuclei interaction such that one can excite a high energy nuclear transition with relatively low energy hard x-ray photons via Doppler-blue-shift. 
Two schemes, the two-color-linear and one-color-cross geometry as showed in Fig.~\ref{NCPT_setups}, have been proposed.
The required  parameters of the used two laser pulses and the nuclear bunch were derived for achieving a complete coherent population transfer between two nuclear ground levels directly (indirectly) via a third level using two $\pi$-pulses (STIRAP) method. We have selected the necessary laser peak intensities with an optimization process, scanning both the laser peak intensity and the time delay between two pulses. An XFELO (SXFEL) laser peak intensity of around 10$^{18}-$10$^{19}$ W/cm$^{2}$ (10$^{20}-$10$^{21}$ W/cm$^{2}$) was found to be sufficient to be used to achieve 100\% nuclear coherent population transfer for the $^{154}$Gd and $^{168}$Er nuclei.
Also,  coherent isomer triggering was considered for $^{97}$Tc and $^{113}$Cd, and the lowest required XFELO and SXFEL peak intensity found are around 10$^{21}$ W/cm$^{2}$ and 3$\times$10$^{23}$ W/cm$^{2}$, respectively, for $^{97}$Tc.  
Moreover, we have derived the first order two-photon resonance condition to connect the error of the XFEL divergence angle and that of a nuclear bunch. This additional condition gives a less strict requirement for the experimental implementation, and can be used to design the parameters of the laser and the nuclear bunches. For estimating the number of the coherently excited nuclei via nuclear coherent population transfer, we have considered the spatial overlap among two XFEL pulses and the LHC nuclear bunch. By using a two-color (one-color) scheme, up to 10$^{5}$ nuclei (30 nuclei) will be coherently excited. Consequently, the two-color scheme is found to be much more efficient for nuclear coherent population transfer.

%%%%%%%%%%%%%%%%%%%%%%%%%%%%%%%%%%%%%%%%%%%%%%%%%%%%%%%%%%%%%%%%%%%%%%%%%%%%%%%%
%%%%%%%%%%%%%%%%%%%%%%%%%%%%%%%%%%%%%%%%%%%%%%%%%%%%%%%%%%%%%%%%%%%%%%%%%%%%%%%%
\end{document}